\documentclass[aps,prb,reprint,amsmath,amssymb,superscriptaddress,longbibliography]{revtex4-2}

\usepackage{graphicx}
\usepackage{dcolumn}
\usepackage{bm}
\usepackage{amsmath}
\usepackage{amssymb}
\usepackage{xcolor}
\usepackage{xspace}
\usepackage{xfrac}
\usepackage{color,soul}
\usepackage{ulem}
\usepackage[colorlinks=true,urlcolor=blue,citecolor=blue,linkcolor=blue,bookmarks=false,pdfstartview={FitH}]{hyperref}

\newcommand{\bk}{{\bf k}}
\newcommand{\tk}{\tilde{\bf k}}

\begin{document}

\title{Josephson effects in twisted nodal superconductors}

\author{Pavel A. Volkov}
\email{pv184@physics.rutgers.edu}
\affiliation{Department of Physics and Astronomy, Center for Materials Theory, Rutgers University, Piscataway, NJ 08854, USA}
\author{S. Y. Frank Zhao} 
\affiliation{Department of Physics, Harvard University, Cambridge, MA 02138, USA}
\author{Nicola Poccia}
\affiliation{Institute for Metallic Materials, IFW Dresden, 01069 Dresden, Germany}
\affiliation{Department of Physics, Harvard University, Cambridge, MA 02138, USA}
\author{Xiaomeng Cui}
\affiliation{Department of Physics, Harvard University, Cambridge, MA 02138, USA}
\author{Philip Kim}
\affiliation{Department of Physics, Harvard University, Cambridge, MA 02138, USA}
\author{J. H. Pixley}
\affiliation{Department of Physics and Astronomy, Center for Materials Theory, Rutgers University, Piscataway, NJ 08854, USA}

\date{August 2021}

\begin{abstract}
    Motivated by the recent proposals for unconventional emergent physics in twisted bilayers of nodal superconductors, we study the peculiarities of the Josephson effect at the twisted interface between $d$-wave superconductors. We demonstrate that for clean  interfaces with a twist angle $\theta_0$ in the range $0^\circ<\theta_0<45^\circ$ the critical current can exhibit nonmonotonic temperature dependence with a maximum at a nonzero temperature as well as a complex dependence on the twist angle at low temperatures. The former is shown to arise quite generically due to the contributions of the momenta around the gap nodes, which are negative for nonzero twist angles. It is demonstrated that these features reflect the geometry of the Fermi surface and are  sensitive to the form of the momentum dependence of the tunneling at the twisted interface. Close to $\theta_0=45^\circ$ we find that the critical current does not vanish due to Cooper pair cotunneling, which leads to a transition to a time-reversal breaking topological superconducting $d+id$ phase. Weak interface roughness, quasiperiodicity, and inhomogeneity broaden the momentum dependence of the interlayer tunneling  leading to a critical current $I_c\sim \cos(2\theta_0)$ with $\cos(6\theta_0)$ corrections. Furthermore, strong disorder at the interface is demonstrated to suppress the time-reversal breaking superconducting phase near $\theta_0=45^\circ$. Last, we provide a comprehensive theoretical analysis of experiments that can reveal the full current-phase relation for twisted superconductors close to $\theta_0=45^\circ$. In particular, we demonstrate the emergence of the Fraunhofer interference pattern near $\theta_0=45^\circ$, while accounting for realistic sample geometries, and show that its temperature dependence can yield unambiguous evidence of Cooper pair cotunneling, necessary for topological superconductivity.
\end{abstract}

\maketitle

\section{Introduction}
Experiments on two dimensional (2D) materials have reached an unprecedented level of control and precision.  2D sheets of atomically thin layers can be  isolated via exfoliation and stacked to make a wide range of devices~\cite{geim2013van}. This approach is expected to be applicable to a variety of materials~\cite{mounet2018two} that can be exfoliated, i.e. have binding energy densities less than $\sim 100$ meV \AA$^{-2}$.
With the development of the ``tear and stack'' approach~\cite{kim2016van} it is now possible to accurately control  the twist angle (to within $\sim 0.1^{\circ}$) between relative sheets of a variety of 2D materials, such as boron nitride~\cite{ni2019soliton,woods2021charge}, 
graphene~\cite{cao2018corr,cao2018sc,yankowitz2019,sharpe2019,Lu2019,jiang2019charge,serlin2020,liu2020,andrei2020graphene}, and transition metal dichalcogenides~\cite{zhang2020flat,regan2020mott,tang2020simulation}. 
The superlattice generated due to the moir\'e pattern that is realized due to the twist, downfolds and strongly renormalizes the single particle spectrum~\cite{lopes2007,bistritzer2011,tarnopol2019,fu2020magic}.
 This approach has successfully led to the observation of correlated insulators and superconductors across a wide range of devices ushering in a new era of ``twistronics''~\cite{carr2017} or ``moir\'e materials''~\cite{balents2020}.
Developing a theoretical description for the resulting single-particle excitations and their instability to interactions has attracted a great deal of theoretical attention \cite{kang2019,Lee2019,repellin2020,vu2021,po2018,hoipo2019,zou2018,vafek2018,guinea2018,senthil2019,tarnopol2019,bernevigmatbg,guinea2020,bernevig-tbg3,bernevig-tbg4}.

The strongly correlated high-temperature cuprate superconductors are layered materials with a highly anisotropic quasi-two-dimensional layered structure \cite{basov2005}. This has recently led to  the realization of atomically thin sheets~\cite{liao2018superconductor,yu2019,frank2019} of Bi$_2$Sr$_2$CaCu$_2$O$_{8+x}$ (BSCCO) with superconducting transition temperatures very close to that measured in bulk samples.
These findings in conjunction with the recent success of moir\'e materials \cite{balents2019} have partly motivated
theoretical proposals to twist nodal superconductors,
at small~\cite{Volkov} and large twist angles~\cite{can2020hightemperature}. 
In the case of singlet $d$-wave superconductors,  
small twist angles $\theta_0 \approx 1^{\circ}$ can possess a magic-angle in the Bogoliubov-de Gennes  spectrum that drives strong interactions between the quasiparticles. Whereas at 
large twist angles (in particular $\theta_0 = 45^{\circ}$) the free energy of the system is lowered by spontaneously generating a phase difference between the two layers, breaking time reversal symmetry (TRSB). 
In this regime, a $d+id$ topological superconducting ground state is realized with a Josephson current-phase ($I-\varphi$) relation between the bilayers that is fundamentally altered~\cite{sigrist1998} from $I\sim  \sin\varphi$ to $I\sim \sin 2\varphi$.

Recent experiments 
on exfoliated thin slabs of BSCCO 
homojunctions with atomically abrupt interfaces
have successfully realized twisted devices with a critical current that strongly depends on the twist angle~\cite{Zhao2021}. 
In these devices, 
BSCCO was cooled to cryogenic temperatures during stacking, which preserved the interfacial structure and superconductivity. In contrast, all the previous experimental attempts to realize superconducting interfaces along the $c$ axis with BSCCO required annealing at high temperature \cite{li1999,takano2002,zhu2021} and yielded varying results on the twist dependence of the critical current. In particular, the strong suppression of the critical current (to zero in the lowest order in tunneling\cite{klemm2001,klemm2005}) has been observed only in one experiment \cite{takano2002}. On the other hand, the cryogenically prepared twist junctions \cite{Zhao2021} demonstrate a dramatic suppression of critical current towards 45$^\circ$, as well as interesting nonmonotonic dependence of the critical current.
Near 45$^\circ$, such twisted junctions exhibit fractional Shapiro steps and a modified Fraunhofer pattern \cite{Zhao2021}, indicating that the Josephson current-phase relation is consistent with the second harmonic, which is required for the topological superconducting ground state \cite{can2020hightemperature}. This experimental breakthrough necessitates the development of a detailed theoretical description of the temperature and twist angle dependence of the critical current as well as the emergent behavior of twisted nodal superconductor interfaces in magnetic fields that goes beyond the previous works \cite{klemm2001,maki2003,klemm2005}. In particular, a possible nonmonotonic temperature dependence of the critical current or signatures of a topological phase near $\theta_0=45^\circ$ including magnetic field effects have not been studied.

In this manuscript we develop the theoretical description of twisted thin slabs of superconductors in terms of their Josephson junction properties. 
Treating the tunnel coupling across the twisted interface as a variable strength Josephson coupling allows for a controlled and systematic perturbative many-body approach. 
This theoretical description was successfully used to describe the experimental data of twisted BSCCO flakes in Ref.~\cite{Zhao2021}.
In the following, we present a detailed derivation of how 
the critical current depends on temperature, twist angle, and magnetic field.
For twist angles close to $\theta_0 =45^{\circ}$ a topological $d+id$ superconducting state is realized with a current-phase relationship that is given by the second harmonic.
For clean and regular interfaces we show that the temperature dependence of the critical current depends sensitively on the Fermi surface geometry and form of the interlayer tunneling, which leads to a nonmonotonic dependence on temperature and twist angle. Presence of nanoscale inhomogeneities at the twist interface washes out these finer details, leading to a critical current that goes like $I_c \sim \cos(2\theta_0)$. For increasing inhomogeneity roughness, we find the topological superconducting phase is destroyed and time reversal symmetry restored.
Using the derived twist angle dependent critical current, the effects of a parallel magnetic field are investigated while incorporating the realistic device geometry used in recent experiments. As a result we are able to demonstrate the emergence of the Fraunhofer pattern of the critical current close to $\theta_0=45^\circ$ and show how it can be used to reveal the current-phase relationship of the twist junction.

The remainder of the paper is organized as follows. In Sec.~\ref{sec:model} we discuss the model investigated and general relations used to compute the current.
In Sec.~\ref{sec:iccoher} we study the effects of Fermi surface geometry and momentum dependent
tunneling with
a clean interface 
where translational symmetry is preserved
and in Sec.~\ref{sec:disorder} we determine how these conclusions are altered by 
considering momentum relaxation due to nanoscale inhomogeneities breaking translational symmetry at the interface. The computed critical current as a function of temperature and twist angle is used to model the twist dependent Josephson coupling to determine the Fraunhofer pattern of the critical current in the presence of a magnetic field in Sec.~\ref{sec:exp}. We conclude in Sec.~\ref{sec:conc}.

\section{Model and general relations}
\label{sec:model}
The recent twisted BSCCO Josephson junction experiments are performed on devices consisting of two flakes of finite thickness, each consisting of a large number of BSCCO unit cells along the $c$ axis. It is established~\cite{Kleiner-1992,Kleiner-1994}, that in bulk BSCCO, the coupling between the superconducting order parameters between the neighboring CuO$_2$ bilayers can be well described by a conventional Josephson coupling.
Consequently, when describing twisted flakes of finite thickness, we will use the effective model of Josephson coupled layers, where coupling between all the layers except at the twisted interface reduces to the conventional Josephson coupling. In the following section we describe the microscopic approach used to compute the interlayer supercurrent across the twisted interface and its dependence on the phase difference of two superconducting bilayers, the temperature, and the twist angle.

Focusing solely on the twisted interface, we start with a model of a superconducting layer with the second layer twisted at an angle $\theta_0$ with respect to the first one. The superconducting layers are described by the Hamiltonian:
\begin{equation}
\begin{gathered}
    \hat{H} =\sum_{\bk,s}
\xi({\bk}) c^\dagger_{\bk s1} c_{\bk s1}
+ 
\xi({\tk}) c^\dagger_{\bk s2} c_{\bk s2}
\\
+\sum_{\bk}
(\Delta({\bk},T) e^{i\varphi} c^\dagger_{\bk\uparrow 1}c^\dagger_{-\bk\downarrow 1}+ \Delta({\tk},T) c^\dagger_{\bk\uparrow 2}c^\dagger_{-\bk\downarrow 2}+ 
{\rm h.c.}),
\end{gathered}
\label{eq:ham0}
\end{equation}
where $\tk = R_{\theta_0}\bk$, $R_{\theta_0}$ being a rotation matrix around the $z$ axis, $\xi(\bk)$ is the single-particle dispersion 
as well as 
$\Delta_1({\bf k},T)=\Delta({\bk},T) e^{i\varphi}$ and 
$\Delta_2(\tilde{\bf k},T)=\Delta(\tilde{\bk},T)$
are the superconducting order parameters in layer one and two respectively, with $\varphi$ being the phase difference between the two.
For the tunneling between the two layers, we assume spin-independent single particle tunneling and follow the approach of Refs. \onlinecite{bistritzer2011,Volkov}, writing the tunneling in momentum space:
\begin{equation}
\hat{H}_{tun} = 
\sum_{\bk,\bk',s} t(\bk,\bk')
c^\dagger_{\bk s1}c_{\bk' s2}
+\mathrm{h.c.}
\label{eq:hamtun}
\end{equation}
where the rotation is accounted for by \eqref{eq:ham0}.
Eq. \eqref{eq:hamtun} represents the most general form of the tunneling Hamiltonian. 
We note that by keeping the tunneling matrix element dependent on momenta in both layers we can consider both the situations where the in-plane momentum is conserved (corresponding to a clean interface) and is not conserved (due to the moire quasiperidocity, roughness, and disorder at the twisted interface).
For a clean system, the tunneling is momentum-conserving 
$t(\bk,\tk')= t(\bk+{\bf G}) \delta_{\bk+{\bf G},\tk'}$,
where ${\bf G}$ is a reciprocal lattice vector \cite{bistritzer2011}. In what follows we will ignore the umklapp processes generated by 
$G\neq0$ in the clean case. 
These processes can be rigorously ignored for a Fermi surface being close to the $\Gamma$ point \cite{Volkov}. While they can be of the same order close to the Brillouin zone edge, we will ignore them here, for qualitative assessment of the tunneling.

To study the Josephson effects in the twisted bilayer we use the general expression for the current-phase relation (CPR) (valid regardless of the tunneling strength)  \cite{golubov2004}
\begin{equation}
    I(\varphi) =
\frac{2e}{\hbar} \frac{d F(T,\varphi)}{d\varphi},
\label{eq:cpr}
\end{equation}
where the free energy is given by
\begin{equation}
      F(T,\theta_0,\varphi)=
    -T{\rm Tr}\log[\hat{G}^{-1}(i\varepsilon_n, \bk) \delta_{\bk,\bk'} - 
    \hat{t}(\bk,\bk')],
    \label{eq:F}
\end{equation}
and
    \begin{equation}
\begin{gathered}
    \hat{G}^{-1}(i\varepsilon_n, \bk)
    = i\varepsilon_n-
    \\
    \begin{bmatrix}
    \xi(\bk)\tau_3+\Delta(\bk)[\cos\varphi \tau_1-\sin\varphi\tau_2]
    & 0\\
    0 &
        \xi(\tk)\tau_3+\Delta(\tk)\tau_1
    \end{bmatrix},
    \\
        \hat{t}(\bk,\bk') 
    =
    \begin{bmatrix}
0
    & t(\bk,\bk')\\
     t^*(\bk,\bk') &
       0
    \end{bmatrix},
\end{gathered}   
\end{equation}
where the matrices act in the layer space. 

For the case of weak tunneling, general expressions can be obtained for $I(\varphi,\theta_0,T)$ 
by expanding the free energy in $t(\bk,\bk')$. The lowest order term reads
\begin{equation}
\begin{gathered}
   I^{(2)}(\varphi,T,\theta_0) = \frac{2e}{\hbar} {\rm Tr}\left[\frac{\partial \hat{G}}{\partial \varphi} \hat{t} \hat{G}\hat{t}\right]=
   \\
   \frac{4e}{\hbar} 
   T\sum_{\varepsilon_n,\bk,\bk'}
   \frac{ |t(\bk,\bk')|^2\Delta(\bk)\Delta(\tk')\sin\varphi}
   {(\varepsilon_n^2+\xi^2(\bk)+\Delta^2(\bk))(\varepsilon_n^2+\xi^2(\tk')+\Delta^2(\tk'))}
   \\
   \equiv I_c^{(2)}(T,\theta_0)\sin\varphi
   \end{gathered}
   \label{eq:i2}
\end{equation}
where $\tk'=R_{\theta_0}\bk'$, and we have introduced  the second order  contribution in $t$ to the critical current $I_c^{(2)}$ via the CPR.
This result for the current  and the CPR corresponds to the conventional linear response obtained from the Kubo formula. 
Importantly, for $\theta_0=45^\circ$, the expression in Eq.~\eqref{eq:i2} vanishes by symmetry for a d-wave superconductor. This can be seen by considering the transformations $x\to-x$ or  $y\to-y$, under these mirror symmetries $\Delta(\tk)$ changes sign, while $\Delta(\bk)$ does not, leading to  $I^{(2)}(\varphi,T,45^\circ) =0$. Note that this statement is still valid for the actual point group of BSCCO crystals \cite{klemm2005}.

The next order in the expansion is given by:
\begin{widetext}
\begin{equation}
\begin{gathered}
   I^{(4)}(\varphi,T,\theta_0) = \frac{2e}{\hbar} {\rm Tr}\left[\frac{\partial \hat{G}}{\partial \varphi} \hat{t} \hat{G}\hat{t}\hat{G}\hat{t}\hat{G}\hat{t}\right]=
   \\
  - \frac{8e}{\hbar}  T 
  \sum_{\varepsilon_n,\bk_1,\tk_2,\bk_3,\tk4}
   \frac{t(\bk_1,\bk_2)t(\bk_2,\bk_3)t(\bk_3,\bk_4)t(\bk_4,\bk_1)\sin\varphi}
   {(\varepsilon_n^2+\xi^2(\bk_1)+\Delta^2(\bk_1))(\varepsilon_n^2+\xi^2(\tk_2)+\Delta^2(\tk_2))}
   \frac{\Delta(\bk_1)\Delta(\tk_2)[\varepsilon_n^2-\xi(\bk_3)\xi(\tk_4)+\Delta(\bk_3)\Delta(\tk_4)\cos\varphi]}
   {(\varepsilon_n^2+\xi^2(\bk_3)+\Delta^2(\bk_3))(\varepsilon_n^2+\xi^2(\tk_4)+\Delta^2(\tk_4))}
   \\
   \equiv I_{1,c}^{(4)}(T,\theta_0)\sin \varphi + I_{2,c}^{(4)}(T,\theta_0)\sin 2\varphi
   \end{gathered}
   \label{eq:i4}
\end{equation}
\end{widetext}
where we assumed time reversal symmetry in the tunneling matrix element ($t(\bk,\bk')=t^*(\bk,\bk')$). 
Two features can be noted in this expression: first, its relative minus sign with respect to Eq. \eqref{eq:i2}. Second, the dependence on the phase difference in Eq.~\eqref{eq:i4} contains both first $\sim \sin\varphi$ and second $\sim 2\sin\varphi\cos\varphi = \sin 2 \varphi$ harmonic dependence on the phase difference, which allows us to define the fourth order contribution in $t$ to the critical current   in the first $I_{1,c}^{(4)}$ and second $I_{2,c}^{(4)}$ harmonic CPRs. Note that the pure first harmonic term has the same properties under mirror symmetries as Eq. \eqref{eq:i2} and hence vanishes exactly at $\theta_0=45^\circ$. On the other hand, close to $T_c$, one observes that the $\sin 2\varphi$ term contains additional square of the order parameter. Consequently, one can expect that at $\theta_0\neq 45^\circ$, the conventional $\sim \sin\varphi$ harmonic will be dominant close to $T_c$.

\subsection{Temperature dependence of the superconducting gap}
\label{sec:deltaT}

To study the temperature dependence of the CPR, the temperature dependence of the gap $\Delta$ has to be included. As we are interested in the qualitative character of this dependence, we will introduce several simplifying assumptions.

Firstly, we assume a weak coupling between the layers, such that the influence of the interlayer hopping \cite{Volkov} and interaction on the magnitude of the mean-field order parameter can be neglected. It follows then that the amplitudes of the order parameters in two layers are independent and equal to each other, i.e. $|\Delta_1|=|\Delta_2|=\Delta$. Note that this does not necessarily imply that the effects of higher-order interlayer tunneling are always negligible for the CPR, and the exact Eqs.~(\ref{eq:cpr}) and (\ref{eq:F}) can be used to study those.

The self-consistency equation for the superconducting gap
within in each layer
then takes the form:
\begin{equation}
\Delta(T,{\bf k}) = T\sum_{\varepsilon_n,{\bf k}'} V_{SC}({\bf k},{\bf k}')
\frac{\Delta(T,\bk')}{\varepsilon_n^2+\xi^2(\bk')+|\Delta(T,\bk')|^2},
\label{eq:MF}
\end{equation}
where $V_{SC}({\bf k},{\bf k}')$ is the intralayer pairing interaction. We will further simplify it by taking an instantaneous interaction with a separable form, i.e. $V_{SC}({\bf k},{\bf k}') = V_{SC} f({\bf k}) f({\bf k}')$, where $f({\bf k})$ vanishes at the nodes. The solutions of \eqref{eq:MF} is then given by:
\begin{equation}
\Delta(T,\bk)=\Delta_0(T) f({\bf k}).
\end{equation}
Finally, we expand $\xi(\bk)$ and $f(\bk)$ in Fourier series in the polar angle in momentum space 
\begin{equation}
    \theta = \arctan(k_y/k_x)
    \label{eqn:polar}
\end{equation} and leave only the lowest harmonics for both. 
In the following manuscript, we will only focus on the case of  a $d$-wave superconductor (that is relevant for twisted BSCCO), in this case we have $\xi(\bk)\to \xi_0(|\bk|)$ and
\begin{equation}
    f(\bk)\to f_0(|\bk|) \cos 2\theta.
    \label{eqn:d-gap}
\end{equation} The integration in \eqref{eq:MF} can be carried out around $k\approx k_F$ such that $ f_0(|\bk|)\approx f_0(k_F)$. One can then define the superconducting gap amplitude at the Fermi level
\begin{equation}
    \Delta(T) \equiv \Delta_0(T) f_0(k_F)
\end{equation}
Using equation \eqref{eq:MF} at $T_c$ to
eliminate
$V_{SC}$ in favor of $T_c$ one arrives at the equation for $\Delta(T)$
\begin{equation}
\begin{gathered}
\sum_{n=-\infty}^{\infty} 
\int_0^{2\pi} d\theta\Big[
\frac{\cos^2 2\theta}{\sqrt{(2n+1)^2+\frac{|\Delta(T)|^2\cos^2 2\theta}{\pi^2T^2}}}
\\
-
\frac{\cos^2 2\theta}{|2n+1|}
\Big]
=0.
\end{gathered}
\label{eq:selfcons}
\end{equation}
In what follows, we use the numerical solution of Eq. \eqref{eq:selfcons} for the temperature dependence of the gap amplitude. For numerical summation here and in what follows, $|n|<|n|_{max} = 20(T_c/T)+50$, which has been checked to be enough for the sum to converge.

\section{$I_c(\theta_0,T)$ for coherent tunnelling}
\label{sec:iccoher}
In this section we demonstrate that momentum-conserving tunneling results in unconventional twist angle- and temperature dependence of the critical current. 
In particular, we show that both the anisotropy of the gap and the Fermi surface result in strong deviations of the low-temperature $I_c(\theta_0)$ from the $\cos 2\theta_0$ form, which is the lowest harmonic consistent with $d$-wave symmetry. These deviations appear much stronger than those observed in recent experiments \cite{Zhao2021} at any temperature. Moreover, the sign-changing nature of the gap is shown to yield a nonmonotonic temperature dependence of $I_c$ at sufficiently large twist angles.

\subsection{Circular Fermi Surface}
\label{sec:iccirc}
\begin{figure}[h]
		\includegraphics[width=\linewidth]{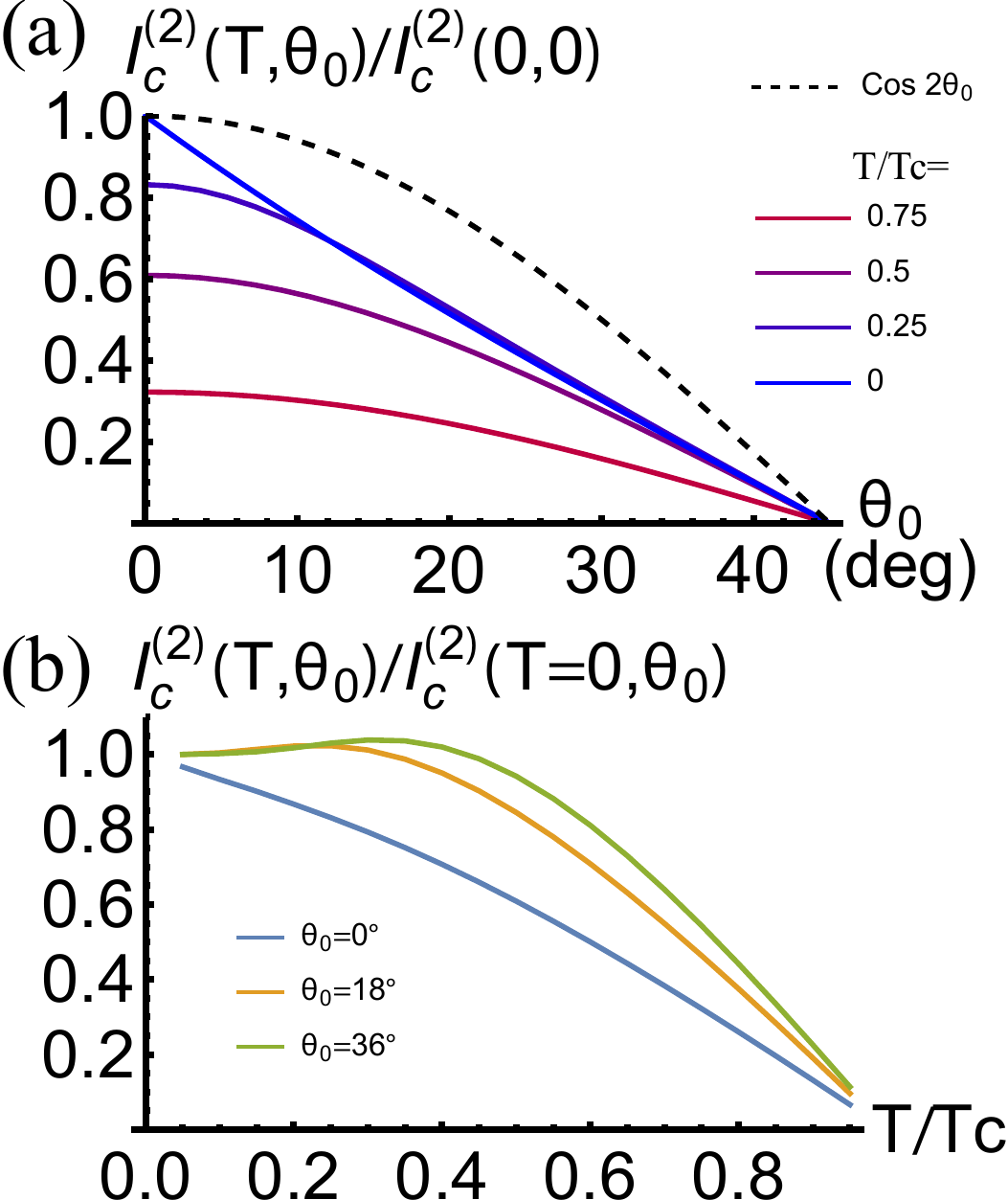}
	\caption{{\bf Second order approximation of the critical current $I^{(2)}_c(T,\theta_0)$ in Eq.~\eqref{eq:i2} for a circular Fermi surface}. (a) Displays the twist angle $\theta_0$ dependence of $I_c^{(2)}$ 
	differing strongly from the $ \cos(2\theta_0)$ behavior for various values of the temperature.   (b) The temperature $T$ dependence of $I_c^{(2)}$ for various twist angles that displays a nonmonotonic behavior.
	}
	\label{fig:i2circ}
\end{figure}

We consider first the simplified model on a circular Fermi surface for  $\xi(\bk)=v_F(k-k_f)$ and a $d$-wave gap symmetry  $\Delta(\bk) = \Delta(T) \cos 2\theta$. We begin with discussing the lowest-order term in the expansion of the CPR in $t$, (\ref{eq:i2}). In Fig. \ref{fig:i2circ} (a) we present the resulting critical current $I_c^{(2)}(T,\theta_0)$ as a function of twist angle for several temperatures. Close to $T_c$, one can expand Eq. \eqref{eq:i2} in the order parameter, resulting in the lowest order in $I_c^{(2)}\sim \int d \theta \Delta(\theta)\Delta(\theta+\theta_0)\sim \cos 2\theta_0$.
However, at low temperatures (Fig. \ref{fig:i2circ} (a)), the twist angle dependence deviates strongly from the $\cos 2\theta_0$ form expected near $T_c$. 
The reason for this deviation are the higher harmonics of $\cos 2\theta_0$ appearing in the denominator of Eq. \eqref{eq:i2} due to the development of an anisotropic $d$-wave gap.

In the region $\theta_0\gtrsim10^\circ$ one also observes that the curve for $T=0$ lies above the one for $T/T_c=0.25$. 
This demonstrates a decrease of $I_c^{(2)}(T,\theta_0)$ on cooling, i.e. a nonmonotonic temperature dependence. We illustrate this in Fig. \ref{fig:i2circ} (b), where indeed $I_c^{(2)}(T,\theta_0)$ has a maximum at an intermediate temperature.

\begin{figure}[h]
		\includegraphics[width=\linewidth]{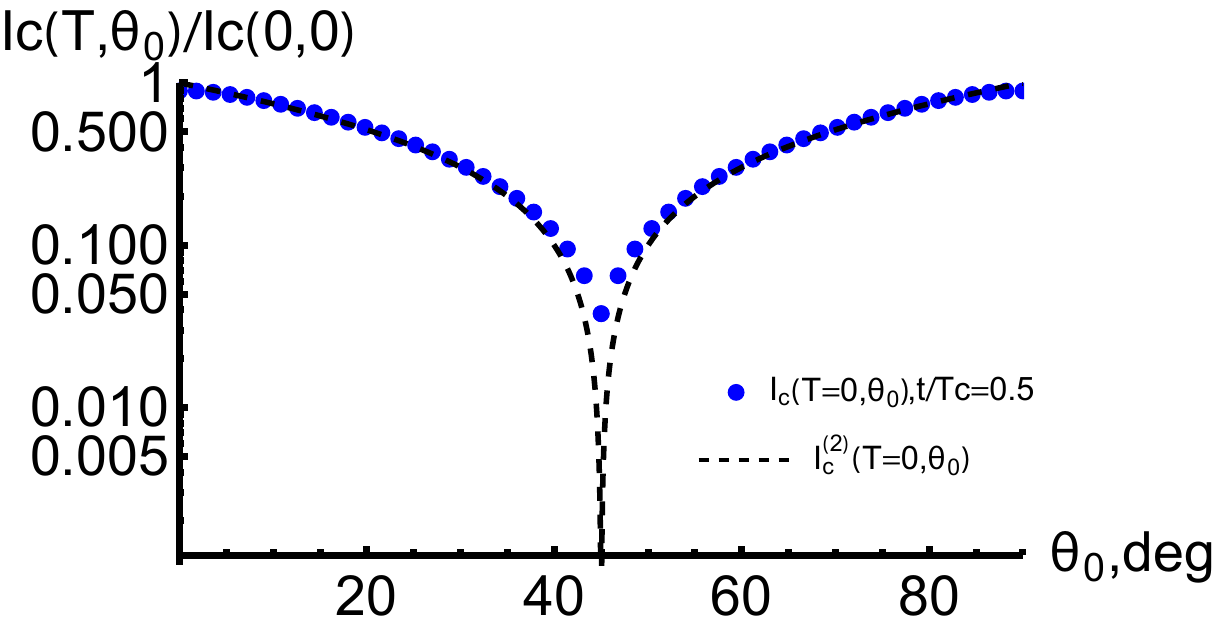}
	\caption{{\bf The twist angle dependence of the normalized critical current defined in Eq.~(\ref{eq:cpr}) for a circular Fermi surface}. Here we take as a representative case $t/T_c=0.5$ and show the data on a log-scale. The dashed line shows the second-order result \eqref{eq:i2} for comparison.
	}
	\label{fig:ic_circ}
\end{figure}

Moving towards the next order in the expansion results, however, in a difficulty. It can be observed that the sums in the perturbative expansion at fourth order in Eq.~\eqref{eq:i4} diverge at $\theta_0=0$ as $\sim 1/T$ for low temperatures at the Dirac nodes $\Delta(\bk)=0$ and $k=k_F$. This suggests that close to $0^{\circ}$ one should use the full expressions in Eqs. (\ref{eq:cpr}) and (\ref{eq:F}) to evaluate the critical current. The tunneling splits two Dirac cones in momentum space at $\theta_0=0$ (due to bonding/antibonding band formation) and away from $\theta_0=0$ the spectrum is gapped for $\varphi\neq0$~\cite{Volkov}, indicating that the divergence is absent in the full formulation. On the other hand, the full current is a rather complicated function of $\varphi$ which has to be maximized to obtain the critical current. Here we take the following approach: away from $\theta_0=0$, we use the expansion in Eqs. (\ref{eq:i2}) and (\ref{eq:i4}) to determine $\varphi_{max}$ and use it in the full expression for the CPR in Eqs. (\ref{eq:cpr}) and (\ref{eq:F}). 
At low twist angles the corrections to the CPR can still be shown to be small for weak tunneling \cite{Volkov}, 
which justifies taking
$\varphi_{max}$ to be equal to $\pi/2$ at low twist angles. Additionally, since the gap opened by the phase difference \cite{Volkov} will generally change the low-temperature behavior of the gap from $T^3$, following from Eq.~\eqref{eq:selfcons} to an exponential one, we focus on the twist angle dependence at low $T$.
In Fig. \ref{fig:ic_circ} we present the twist angle dependence of the critical current compared to the second-order expansion result for $t/T_c=0.5$. Away from $\theta_0=45^\circ$ one observes almost no difference between the two, suggesting that the second-order expansion constitutes a good approximation. However, while Eq.~\eqref{eq:i2} manifestly goes to zero at $\theta_0=45^\circ$ by symmetry, the full critical current does not. This yields a qualitative explanation of the observation of a nonzero critical current at $\theta_0=45^\circ$ in otherwise strongly angle-dependent results of Ref. \onlinecite{Zhao2021}.

\subsection{Cuprate-like Fermi surface}
\label{sec:iccupr}
We now consider the qualitative effects of a non-circular Fermi surface. In particular, we take a Fermi surface appropriate for cuprates that can be deduced from the tight-binding model dispersion for a square lattice:
\begin{equation}
\begin{gathered}
\xi(\bk)
=
-2 t_0 (\cos k_x +\cos k_y)
- 4 t_0' \cos k_x \cos k_y 
\\
- 2 t_0'' (\cos 2 k_x +\cos2k_y)
-\mu,
\end{gathered}
\label{eq:tb}
\end{equation}
and a $d$-wave superconducting gap  on the square lattice
\begin{equation}
    \Delta(\bk) = \Delta(T) (\cos k_x -\cos k_y).
\end{equation}
We use the parameters appropriate for BSCCO \cite{markiewicz2005}: $t_0= 126$ mev, $t_0'=-36$ mev, $t_0''=15$ mev, $\mu=15$ meV and take $k_B T_c=9$ meV (corresponding to $T_c\approx 90$ K). Note that the unit cell of BSCCO contains two CuO$_2$ layers; we ignore this bilayer structure as we study here the qualitative behavior of $I_c(\theta_0,T)$. In this subsection we keep the tunneling to be momentum independent but generalize this  below. For numerical calculation in this and next section we additionally rotated the momenta by $-\theta_0/2$.

\begin{figure}[h]
		\includegraphics[width=\linewidth]{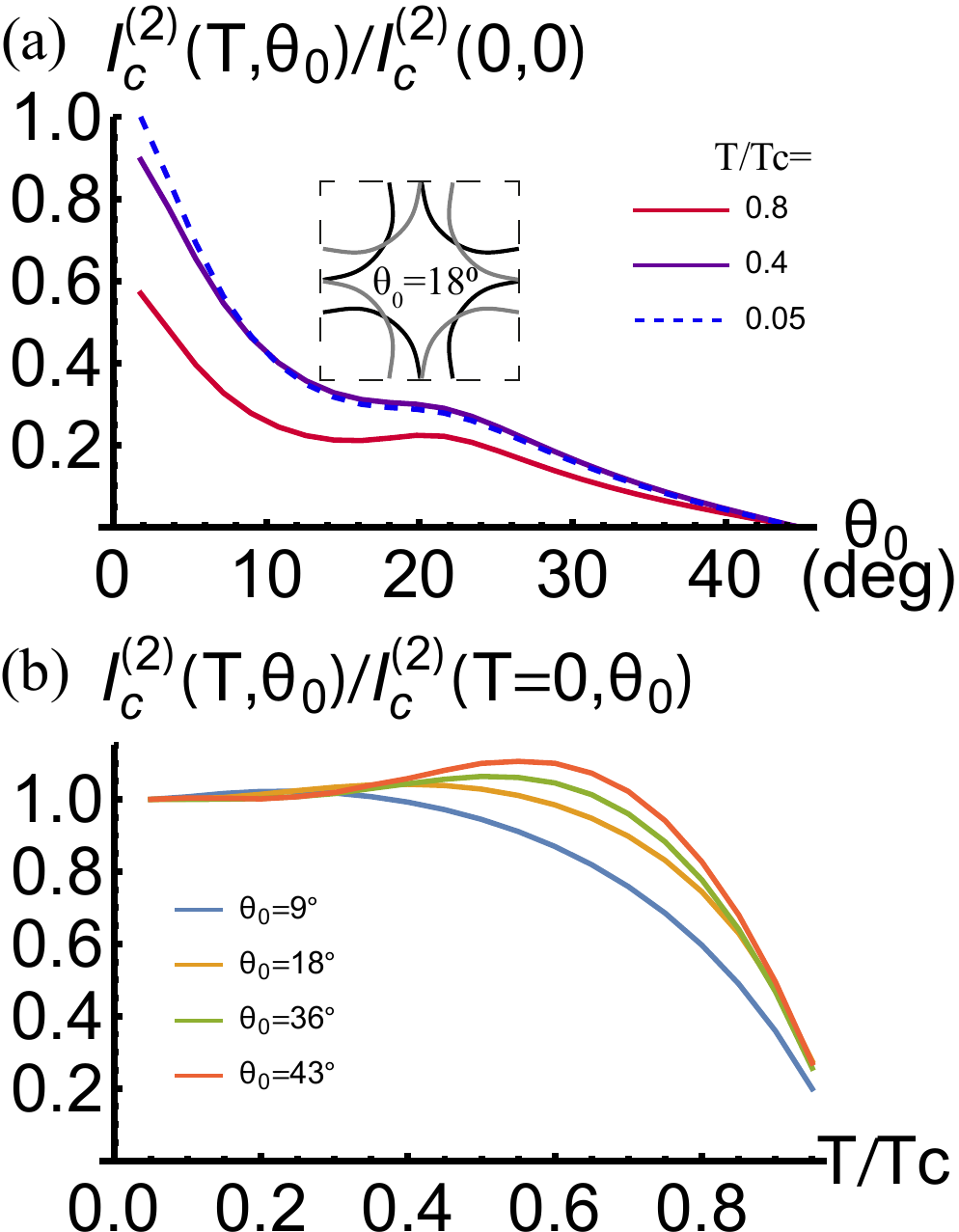}
	\caption{
	{\bf Second order approximation of the critical current $I^{(2)}_c(T,\theta_0)$ in Eq.~\eqref{eq:i2} for a cuprate-like Fermi surface with parameters from BSCCO}.
	(a) Twist angle ($\theta_0$) dependence of $I^{(2)}_c$ for various temperatures ($T$) displaying a local maximum near $\theta_0\approx 20^{\circ}$. Inset shows an overlay of two twisted cuprate-like Fermi surfaces at $\theta_0=18^\circ$ corresponding to a broad maximum in $I_c^{(2)}$ at low temperatures. (b) Nonmonotonic temperature dependence of $I_c^{(2)}$, which becomes more pronounced near $\theta_0=45^\circ$.}
	\label{fig:i2cupr}
\end{figure}

In Fig. \ref{fig:i2cupr} we show the twist angle and temperature dependence of $I_c^{(2)}(\theta_0,T)$. One observes very pronounced deviation from the $\cos 2\theta_0$ form. In particular, the steep initial decrease of $I_c^{(2)}(\theta_0,T\ll T_c)$ with $\theta_0$ resembles the results of experiments on whisker twist junctions \cite{takano2002}. We note, that unlike Ref. \cite{maki2003}, the deviation from the $\cos 2\theta_0$ form appears already in the lowest-order tunneling approximation, consistent with previous works \cite{klemm2005}. Another feature that is present in our results is a broad maximum in $I_c$ at around $\theta_0=20^\circ$. As  shown in the inset of Fig. \ref{fig:i2cupr} (a), close to this twist angle,  the Fermi surfaces of the two layers start crossing each other near the Brillouin zone boundary. The contribution of this region to Eq. \eqref{eq:i2} is positive and is maximized when the Fermi surfaces cross (i.e. $\xi(\bk)=\xi(\tk)=0$), suggesting that the maximum reflects the appearance of this crossing. A more quantitative discussion of this point is presented in Sec. \ref{sec:icqualit}.

Moreover, we find a nonmonotonic temperature dependence of $I_c$ (Fig. \ref{fig:i2cupr} (b)), that becomes relatively more pronounced towards $\theta_0=45^\circ$ (we note however that sufficiently close to $\theta_0=45^\circ$ the higher-order terms in $t$ will become dominant). The nonmonotonicity in this case appears stronger than for the circular Fermi surface case.

\subsubsection{Momentum-dependent tunneling}

Finally, we address the effects of the momentum dependence of the tunneling. This is indeed relevant for cuprates, where the dominant tunneling between the $d_{x^2-y^2}$-like orbitals occurs via  intermediate $s$-like orbitals  \cite{andersen1995}, leading to $t(\bk) = t_z (\cos k_x-\cos k_y)^2$ in the bulk of the material. At a twisted interface, $d_{x^2-y^2}$-like orbitals in the twisted layer are rotated leading to:
\begin{equation}
    t(\bk) = t_z (\cos k_x-\cos k_y)(\cos \tilde{k}_x-\cos \tilde{k}_y),
    \label{eq:t(k)}
\end{equation}

\begin{figure}[h]
		\includegraphics[width=\linewidth]{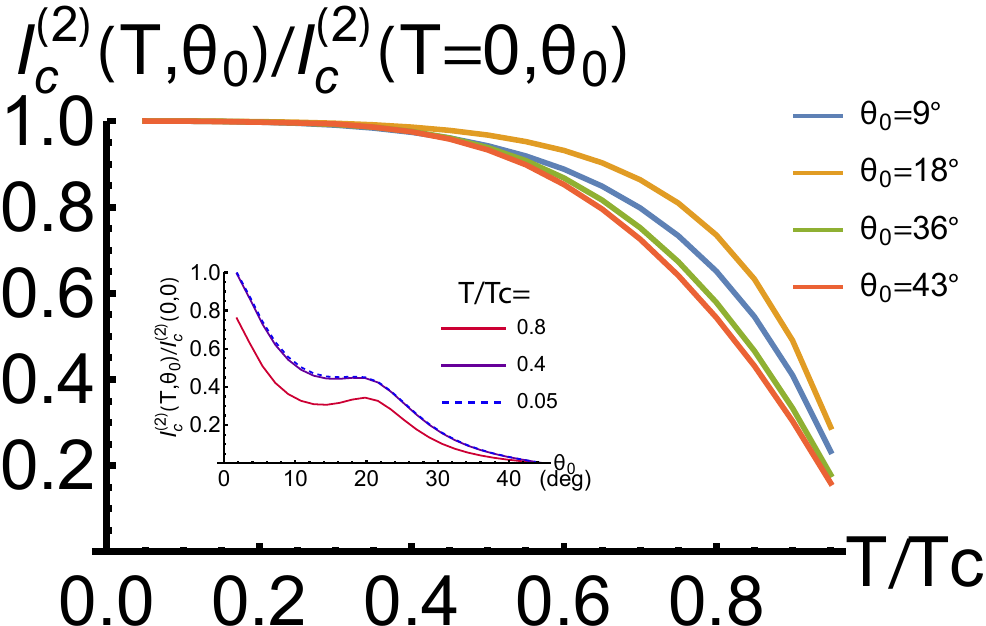}
	\caption{
	{\bf Second order approximation of the critical current $I^{(2)}_c(T,\theta_0)$ in Eq.~\eqref{eq:i2} incorporating the momentum dependent tunneling relevant for BSCCO in Eq.~\eqref{eq:t(k)}}. As in Fig.~\ref{fig:i2cupr} we are also taking a cuprate-like Fermi surface with parameters from BSCCO. The temperature dependence of the normalized $I_c^{(2)}$ no longer displays the maximum found in Fig.~\ref{fig:i2cupr}(b). Inset shows the twist angle dependence of $I_c^{(2)}$ that remains qualitatively similar to \ref{fig:i2cupr} (a).}
	\label{fig:i2tk}
\end{figure}

In Fig. \ref{fig:i2tk} we present the $I_c^{(2)}(\theta_0,T)$ computed with the momentum-dependent tunneling \eqref{eq:t(k)}. Remarkably, the temperature dependence of $I_c^{(2)}$ is always monotonic in this case, while the twist angle dependence is quite similar to the case of a momentum-independent tunneling. Thus, we see that the temperature and twist angle dependence of the critical current for a twist junction depends strongly on the Fermi surface geometry and form of the tunneling in the coherent (momentum-conserving) tunneling limit. We note that strong deviations from $I_c(\theta_0)$ going like $\sim \cos 2 \theta_0$ 
is observed for all models considered, which is in contrast to the resent experiments on cryogenically prepared twist junctions \cite{Zhao2021}.
As we show in Sec.~\ref{sec:disorder} below, in the presence of weak momentum relaxing effects at the twist junction, the $\cos 2\theta_0$ dependence  appears clearly.

\subsection{Qualitative assessment of $I_c(\theta_0,T)$: nodal/antinodal dichotomy}
\label{sec:icqualit}
We now present qualitative arguments allowing additional insight into the results of the previous sections. Let us start with the non-monotonic temperature dependence of Sec. \ref{sec:iccirc} and \ref{sec:iccupr}. As the gap amplitude, per Eq. \eqref{eq:MF} is strictly monotonic function of temperature, one expects that if the summand in Eq. \eqref{eq:i2} was positive for all $\bk$, the resulting $I_c(T)$ would be monotonic. This is however, not the case at a finite twist angle. In particular, in between two nodal lines of the superconducting gap, that were aligned at $\theta_0=0$, the order parameter has different sign for two layers, leading to a negative contribution to Eq. \eqref{eq:i2}.

\begin{figure}[h]
		\includegraphics[width=\linewidth]{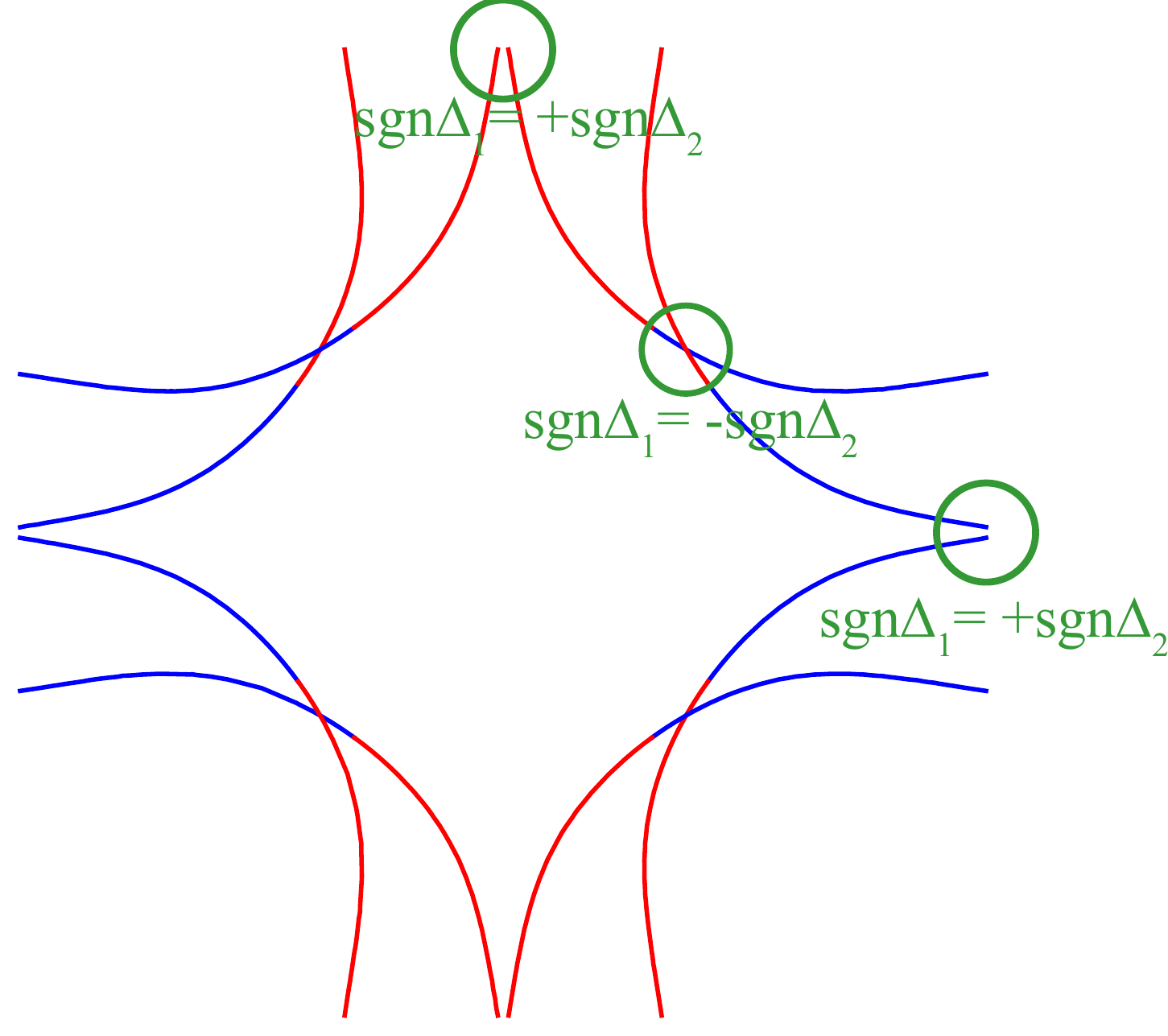}
	\caption{
	{\bf Twisted cuprate like Fermi surfaces.}
	Fermi surface schematic for $\theta_0=18^\circ$ with the order parameter sign shown by color. In the nodal region, the order parameters with overlapping momenta have opposite signs due to twist.
	}
	\label{fig:FSschem}
\end{figure}

This is especially clear in the case of a cuprate-like Fermi surface at a finite twist angle, which is displayed in Fig. \ref{fig:FSschem}. As has been noted above, the contribution to Eq. \eqref{eq:i2} is enhanced near the points where Fermi surfaces cross. At low twist angle, such a crossing occurs in the near-nodal (N) region, where the order parameter has opposite sign for two layers. For larger twist angles, an additional crossing appears in the antinodal (AN) region (close to the Brillouin zone boundary). There, on the contrary, the order parameters of the two layers have the same sign. The contributions of these regions to $I_c$ reads:
	\begin{equation}
	\begin{gathered}
	\delta I_c^N (T) \sim
	\frac{\Delta_N(\theta_0,T) \tanh\frac{\Delta_N(\theta_0,T)}{2 T}}{|{\bf v}_F^{1,N}\times {\bf v}_F^{2,N}|},
	\\
	\delta I_c^{AN}(T) \sim 
	\frac{\Delta_{AN}(\theta_0,T) \tanh\frac{\Delta_{AN}(\theta_0,T)}{2 T}}{|{\bf v}_F^{1,AN}\times {\bf v}_F^{2,AN}|},
	\end{gathered}
	\end{equation}
where ${\bf v}_F^{(1,2),(N,AN)}$ are the Fermi velocities at the points where the Fermi surfaces cross and $\Delta_{N,AN}$ are the gaps in the N and AN regions. At low twist angles, only the negative nodal contribution is relevant. It is negative and becomes larger in magnitude on cooling, providing an explanation for the decreasing $I_c$. Its magnitude is suppressed at low twist angles due to the smallness of the gap at the Fermi surface crossing $\Delta_N(\theta_0,T)\sim \theta_0$. This explains why the nonmonotonicity is enhanced by twist.

At larger twist angles, the antinodal crossing appears, which contributes an enhanced positive correction to $I_c$. This is indeed what is seen to occur in Fig. \ref{fig:i2cupr} (a). At low temperatures, both nodal and antinodal contributions saturate to finite values. However, $\Delta_{AN}(T)/\Delta_{N}(T)\sim \mathrm{const.}>1$; consequently, for temperatures $2\Delta_{N}(0)<T<2\Delta_{AN}(T=0)$,  the $\tanh$ in $\delta I_c^{AN}(T)$  is already saturated to a constant, while the $\tanh$ in  $\delta I_c^{N}(T)$ will continue to grow in absolute magnitude on cooling. Thus for temperatures $2\Delta_{N}(0)<T<2\Delta_{AN}(0)$, the total $\delta I_c^{AN}(T)+\delta I_c^{N}(T)$ will decrease on cooling, implying a nonmonotonic $I_c(T)$.

Finally, the effect of the momentum dependent tunneling on the temperature dependence of $I_c$ can be understood from this picture. Indeed, the tunneling Eq. \eqref{eq:t(k)} is very strongly suppressed in the nodal region, vanishing as $\theta_0^2$ for low twist angles. This suppresses the contribution of the nodal region to $I_c$ in agreement with it being the source of nonmonotonicity.

\subsection{$I_c(T)$ at $\theta_0=45^\circ$ due to cotunneling}

As has been shown above (see, e.g., Fig. \ref{fig:ic_circ}), the second-order tunneling in Eq.~\eqref{eq:i2} dominates the Josephson effect apart from in the vicinity of $\theta_0=45^\circ$, where the cotunneling of Cooper pairs in Eq.~\eqref{eq:i4} takes over. We now consider the temperature dependence of the cotunneling critical current.

\begin{figure}[h]
		\includegraphics[width=\linewidth]{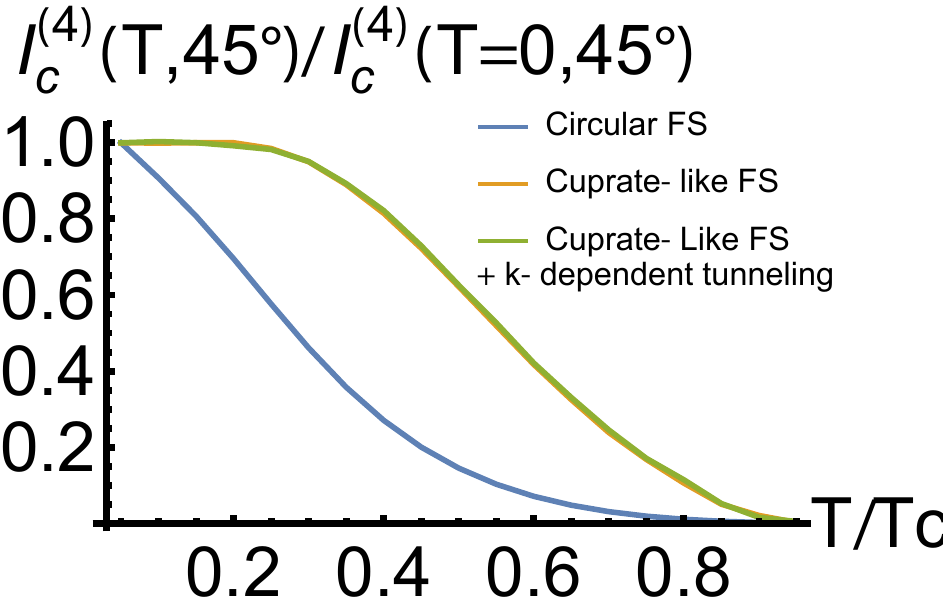}
	\caption{
	{\bf The fourth order contribution to the critical current  at $\theta_0=45^{\circ}$.}
	Temperature dependence of the cotunneling critical current, Eq. \eqref{eq:i4}, for three models with coherent tunneling: circular FS (blue), cuprate-like FS (yellow) and  cuprate-like FS with momentum-dependent tunneling (green); the latter two are almost identical.
	}
	\label{fig:i4coh}
\end{figure}

In Fig. \ref{fig:i4coh} we present the temperature dependence of the cotunneling critical current for the three models considered above. In all of the cases, the dependence is more steep, than for the tunneling critical current (see Fig. \ref{fig:i2circ} (b), Fig. \ref{fig:i2cupr} (b) and Fig. \ref{fig:i2tk}). On approach to $T_c$, the cotunneling critical current is suppressed much stronger, than the tunneling one, as is expected from the general expression in Eq.~\eqref{eq:i4}. This distinct temperature dependence may serve as a qualitative indicator of the presence of a second harmonic in the CPR.

\section{Effects of interface inhomogeneity on $I_c(\theta_0,T)$}
\label{sec:disorder}

Here we study the consequences of the broken translational symmetry at the interface due to lattice supermodulations, moir\'e quasi-periodicity, atomic scale interface roughness, or disorder all of which  result in the in-plane momentum  not being conserved during tunneling $t(\bk,\bk'\neq\bk)\neq0$. We will work in the weak tunneling approximation here, using the expansion in Eqs.~(\ref{eq:i2}) and (\ref{eq:i4}). Several models for $t(\bk,\bk')$ can be considered. First, for a purely incoherent tunneling $t(\bk,\bk')=t_0$, corresponding to atomic-scale disorder, such as in the case of the Ambegaokar-Baratoff formula for $s$-wave superconductors \cite{AB,ABerr}, the Eqs. (\ref{eq:i2}) and (\ref{eq:i4}) yield {\it identically zero} due to the $d$-wave symmetry of the order parameters. For a superposition of fully coherent and incoherent terms $t(\bk,\bk')=t_0+t_1 \delta_{\bk,\bk'}$ it is evident, that only $t_1$ will contribute in the lowest order in Eq.~(\ref{eq:i2}). In the recent experiments \cite{Zhao2021}, the critical current at the interfaces prepared at $\theta_0=0$ has been observed to be similar to the one expected between individual layers in the bulk. That rules out the presence of strong atomic-scale disorder at the twist interface.

For the more realistic case of weak nanoscale disorder (such as structural supermodulations \cite{Poccia2020}), with a length scale significantly larger than the unit cell size, the tunneling has a characteristic momentum spread that is smaller than the size of the Brillouin zone. We consider the case where tunneling is not exactly momentum conserving, modeled with a spread of $\sigma$ in typical momentum differences $|{\bf k} - {\bf k}'|$.
This can be implemented by replacing $|t(\bk,\bk')|^2$ 
with a function with a width $\sigma$ (e.g. a Gaussian)
denoted $|t_{\sigma}(\bk,\bk')|^2$ \cite{klemm2001,klemm2005}. We choose the normalization such that in the limit $\sigma \to 0$ we recover coherent tunneling, i.e.
$|t(\bk,\bk')|^2= \delta({\bf k}-{\bf k}')$,
i.e.
\begin{equation}
    |t_{\sigma}(\bk,\bk')|^2=\frac{t_0^2}{2 \pi \sigma^2} e^{-\frac{|{\bf k} - {\bf k}'|^2}{2\sigma^2}}.
    \label{eqn:tk}
\end{equation}
For momenta close to the Fermi surface one can further split the constraint on the tunneling momentum into those on the momentum magnitude and the polar angle in Eq.~\eqref{eqn:polar}:
\begin{equation}
\begin{gathered}
|{\bf k} - {\bf k}'|^2 
=(k-k')^2+4kk'\sin^2\frac{\theta-\theta'}{2}
\\
\approx (k-k')^2 + k_F^2(\theta-\theta')^2,
\end{gathered}
\end{equation}
where $\theta^{(')}+2\pi \equiv \theta^{(')}$. It follows then that the angular spread of the tunneling is equal to $\tilde{\sigma} = \sigma/k_F$.

\subsection{Second order tunneling $I^{(2)}$}
First, we consider the second order tunneling process with interfacial disorder at the twist junction.
The angular integrals of $\bk, \bk'$ in Eq.~(\ref{eq:i2}) are performed in Appendix~\ref{sec:appendixA} using a Fourier expansion.
To make further progress analytically, we take $\xi(\bk)=\xi(k)$ and $\Delta(\bk) = \Delta(T)\cos 2\theta $ as in section \ref{sec:deltaT} and limit ourselves to the lowest terms in the Fourier series.

\begin{figure}[h!]
		\includegraphics[width=\linewidth]{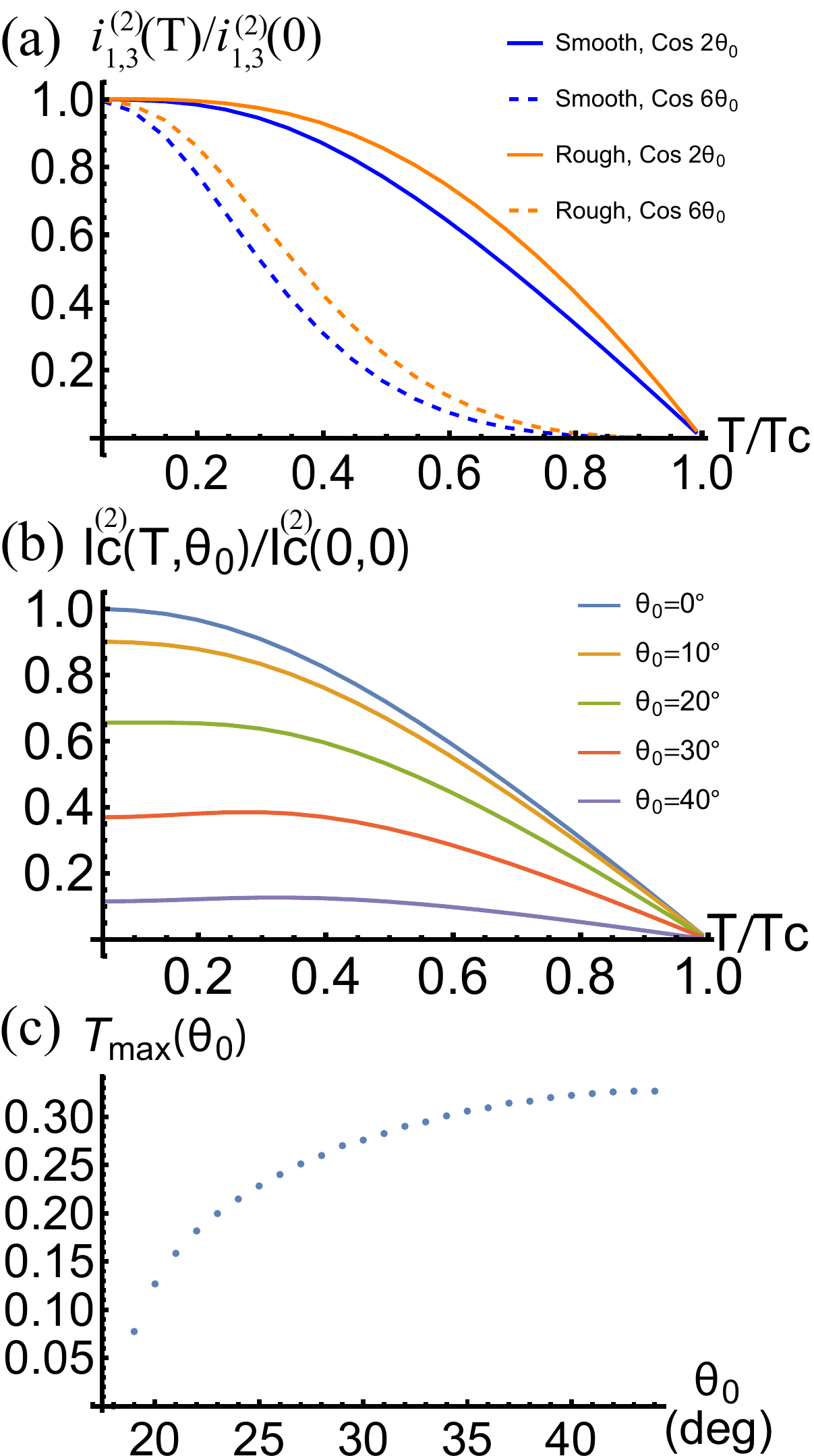}
	\caption{
	{\bf Effects of weak interfacial inhomogeneities.}
	(a) Temperature dependence of the normalized tunneling critical current harmonics normalized to their values at $T=0$ for smooth ($i^{(2)}_1(T)/i^{(2)}_1(0),i^{(2)}_3(T)/i^{(2)}_3(0)$ in \eqref{eq:ictclean}) (blue, "smooth") and rough ($i^{(2)'}_1(T)/i^{(2)'}_1(0),i^{(2)'}_3(T)/i^{(2)'}_3(0)$ in \eqref{eq:ictdirty} ) (yellow, "rough") interface inhomogeneity compared to the superconducting coherence length. 
	(b) Temperature dependence of the second-order critical current $I_c^{(2)}(T,\theta_0)$ (normalized to its $T=0,\theta_0=0$ value) for several values of the twist angle with two harmonics included in the smooth interface limit \eqref{eq:ictclean} with $\tilde{\sigma}=0.15$.
	(c) Temperature $T_{max}(\theta_0)$ of the critical current maximum in (b).
	}
	\label{fig:icdisor}
\end{figure}

For the integral over the magnitude of $\bk, \bk'$ (i.e. $k,k'$) in 
Eq.~(\ref{eq:i2}),
we consider two limiting cases. 
For smooth inhomogeneity (SI) we assume $\hbar v_F \sigma \ll \sqrt{\Delta^2(T)+(\pi T)^2}$ that can be valid at all $T$. This limit corresponds to the inhomogeneity length scale being longer than the BCS coherence length of the superconductor.
We can further simplify the result by taking the limit $\sigma\to 0$ in the $k,k'$ integral 
(see Eq.~\eqref{eq:i2incoh}) 
to obtain
\begin{equation}
\begin{gathered}
I^{(2)}_{\mathrm{SI}}(\varphi,T,\theta_0) \approx
\mathcal{A}t_0^2
e^{-2\tilde{\sigma}^2}
(i^{(2)}_1(T)\cos(2\theta_0)
\\
+i_3(T)\cos(6\theta_0)e^{-16\tilde{\sigma}^2})\sin\varphi,
\end{gathered}
\label{eq:ictclean}
\end{equation}
where for simplicity of presentation we have defined  the constant $\mathcal{A}=\frac{e   k_F} {4 \pi^3 \hbar^2 v_F}$, and introduced the contributions to the first $i_1^{(2)}(T)$ and third harmonics $i_3^{(2)}(T)$ of the Fourier expansion that are evaluated in Appendix~\ref{sec:appendixA}. In the limiting cases of $T\approx 0$ and $T_c$  we obtain
\begin{equation}
    i^{(2)}_1(T)
\approx
\begin{cases}
2(\log 4 -1) &T\to 0\\
0.1 \frac{\Delta^2(T)}{T_c^2} &T\to T_c
\end{cases}
\end{equation}
as well as
\begin{equation}
   i^{(2)}_3(T)
\approx
\begin{cases}
2(\log 4 -4/3) &T\to 0\\
4\cdot10^{-5}\frac{\Delta^6(T)}{ T_c^6} &T\to T_c.
\end{cases} 
\end{equation}
Note that $I_c(0)$ is then independent of $\Delta(T)$. This suggests that qualitative signature of this regime is the independence of $I_c(T=0)$ of $T_c$, the latter being controlled, by, e.g., doping.
	This is also consistent with the form $I_c(\theta_0)\sim \cos(2\theta_0)$ as observed in experiments on BSCCO twist junctions \cite{Zhao2021}.

In the opposite limit of rough inhomogeneity (RI) $\hbar v_F \sigma \gg \sqrt{\Delta^2(T)+(\pi T)^2}$
we get the more usual Ambegaokar-Baratoff \cite{ABerr} like expression
(see Appendix~\ref{sec:appendixA})
\begin{equation}
\begin{gathered}
I^{(2)}_{\mathrm{RI}}(\varphi,T,\theta_0) \approx
\frac{\mathcal{A}t_0^2
e^{-2\tilde{\sigma}^2}
 }{\sqrt{2\pi}\hbar v_F \sigma}\times
\\
\left(\tilde{i}^{(2)}_1(T)\cos(2\theta_0)
+\tilde{i}^{(2)}_3(T)\cos(6\theta_0)e^{-16\tilde{\sigma}^2}\right)\sin\varphi,
\end{gathered}
\label{eq:ictdirty}
\end{equation}
where the first harmonic is now
\begin{equation}
\tilde{i}^{(2)}_1(T)  
\approx
\begin{cases}
6.035 \Delta(T) &T\to 0\\
\frac{\pi^2}{4} \frac{\Delta^2(T)}{T_c} &T\to T_c
\end{cases}
\end{equation}
and the third harmonic is given by
\begin{equation}
\tilde{i}^{(2)}_3(T) 
\approx
\begin{cases}
0.18 \Delta(T) &T\to 0\\
3\cdot10^{-4}\frac{\Delta^2(T)}{T_c} &T\to T_c.
\end{cases}
\end{equation}
Again, we use the constant $\mathcal{A}=\frac{e   k_F} {4 \pi^3 \hbar^2 v_F}$, and have introduced distinct  contributions to the first $\tilde{i}_1^{(2)}(T)$ and third harmonics $\tilde{i}_3^{(2)}(T)$ of the Fourier expansion in the limit of rough inhomogeneity at the twist interface. Note that the distinction "rough" does not imply a strong disorder at the interface, but rather characterizes the length scale of the typical inhomogeneities.

Let us now consider the temperature dependence of the lowest-order critical current following the $\cos(2\theta_0)$ twist angle dependence. In Fig. \ref{fig:icdisor} (a) we present the temperature dependence of the lowest twist-angle harmonic of the critical current deduced from (\ref{eq:ictclean} ,\ref{eq:ictdirty}) and taking the temperature dependence of the gap from the numerical solution of Eq. \eqref{eq:selfcons}. Importantly, in both cases it appears monotonic. These results suggest that at the level of weak tunneling, the nonmonotonic temperature dependence of $I_c$ is intimately related to coherence of the tunneling.

In both cases of smooth and rough inhomogeneity the $\cos(6\theta_0)$ contribution appears to be strongly suppressed numerically (in addition to the exponential suppression due to angular spread): by an almost order of magnitude at low $T$ and by several orders of magnitude close to $T_c$. 
The $\cos(6\theta_0)$ contribution has the same sign as the $\cos(2\theta_0)$ one in both the clean and rough limit. However, the relative sign of the two contributions changes with $\theta_0$ well before $\theta_0=45^\circ$. In the clean limit, this leads to a clear nonmonotonic temperature dependence of $I_c$ (Fig. \ref{fig:icdisor} (b)), which shows a maximum at a finite temperature for $\theta_0>18^\circ$, close to the values observed in experiment \cite{Zhao2021}.

To conclude this subsection, we have found that relaxing momentum conservation at the twist interface naturally accounts for the observation of 
\begin{equation}
    I_c(\theta_0,T)\sim \cos (2\theta_0)
\end{equation}
at low temperatures seen in recent experiments~\cite{Zhao2021}. The high value of the critical current of the twist junction observed in experiment~\cite{Zhao2021} also indicates that the momentum relaxation is arising from nanoscale inhomogeneities, such as ones that arise from structural supermodulation \cite{Poccia2020}, and not atomic-scale disorder. This is consistent with the atomically sharp interfaces with structural supermodulations observed using transmission electron microscopy in Ref.~\cite{Zhao2021}.

\subsection{Fourth order tunneling $I^{(4)}$}
\begin{figure}[h]
		\includegraphics[width=\linewidth]{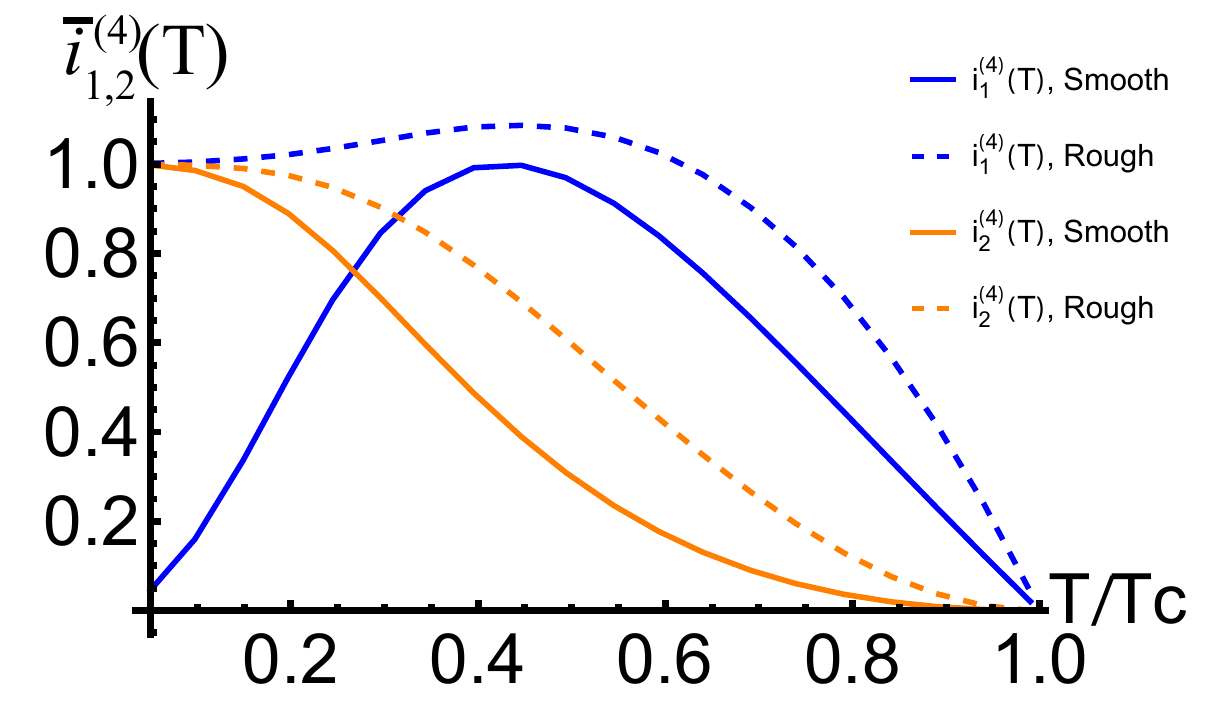}
	\caption{
	{\bf The fourth order contribution to the critical current  with weak interfacial inhomogeneities.}
	Temperature dependence of the normalized fourth-order critical current harmonics $i^{(4)}_{1,2}(T)$, Eq. \eqref{eq:ict4disor2}. $i^{(4)}_{1}(T)$ in the weak roughness limit is normalized to its maximal value, while the rest - to their values at $T=0$.
	}
	\label{fig:ic4disor}
\end{figure}

Finally, we discuss the fourth-order tunneling contribution to the critical current. Applying the same expansion in twist angle harmonics to  Eq. \eqref{eq:i4}, we obtain the following result for the two leading  harmonics with a common form to both the SI and RI regimes
\begin{equation}
\begin{gathered}
   I^{(4)} = I^{(4)}_{1,c}(T) \cos 2\theta_0 \sin\varphi+I^{(4)}_{2,c}(T,\theta_0)\sin 2\varphi 
   \end{gathered}
   \label{eq:ict4disor1}
\end{equation}
where the coefficients of the first harmonic $I^{(4)}_{1,c}(T)$ and second harmonic $I^{(4)}_{2,c}(T,\theta_0)$ of the CPR are given by
\begin{equation}
\begin{gathered}
   I^{(4)}_{1,c}(T) =   - \frac{e t_0^4 k_F \sigma^2 e^{-3\tilde{\sigma}^2}}{\hbar^2 v_F (2\pi)^6} 
   i^{(4)}_1(T).
   \\
  I^{(4)}_{2,c}(T,\theta_0)
  =
   - \frac{e t_0^4 k_F \sigma^2 e^{-4\tilde{\sigma}^2}}{\hbar^2 v_F (2\pi)^7} 
   (\cos 4\theta_0+2 e^{-4\tilde{\sigma}^2})
   i^{(4)}_2(T)
   \end{gathered}
   \label{eq:ict4disor2}
\end{equation}
and the expressions for $i_1^{(4)}(T)$ and $i_2^{(4)}(T)$ are given in Appendix~\ref{sec:appendixA} in both the SI and RI regimes; their temperature dependence is shown in Fig. \ref{fig:ic4disor}. We find that in the limit of SI $i_1^{(4)}$ is strongly suppressed at low temperatures in contrast to $i_2^{(4)}$, which saturates to a non-zero value. Whereas in the opposing limit of a RI we find both contributions survive to low temperatures.

Several qualitative conclusions can be drawn from \eqref{eq:ict4disor1}. First, the suppression of this term with roughness is much stronger than for the usual tunneling term, due to the additional factors of $\sigma^2e^{-\tilde{\sigma}^2}$. Additionally, the twist angle dependence of the second-harmonic CPR ($\sin2\varphi$) term 
is modified due to disorder via the term
$\cos 4\theta_0 + 2 e^{-4\tilde{\sigma}^2}$ in Eq.~\eqref{eq:ict4disor2}. This implies profound consequences for the system close to $\theta_0=45^\circ$: if the disorder strength is sufficiently large, it is possible to destroy the topological superconducting phase at $\theta_0=45^{\circ}$ because the $\cos 4\theta_0$ will then be dominant and negative, which will change the overall sign of the second harmonic in the CPR. In that case, the state with a dominant second harmonic in the CPR would still have a free energy minimum at $\varphi=0$, indicating the absence of a spontaneous time reversal symmetry breaking, and, consequently, the destruction of the topological phase. For the Gaussian momentum smearing used here in Eq.~\eqref{eqn:tk},  we find a  topological superconductor to trivial superconductor transition occurs at a critical value of the disorder strength $\tilde{\sigma}_c\approx0.42$, which corresponds to an angular spread  of around $\pm24^\circ$ (see Eq.~\eqref{eqn:tk}) for incoherent tunneling. Importantly, such a broad interlayer momentum dependent tunneling is inconsistent with the atomically sharp interface observed experimentally~\cite{Zhao2021}.
In summary,  for a twist $\theta=45^{\circ}$ and $\tilde \sigma<\tilde{\sigma}_c$ the ground state is a topological superconductor that breaks time reversal symmetry, whereas for $\tilde \sigma>\tilde{\sigma}_c$ the superconductor is trivial and the time reversal symmetry is restored by inhomogeneity.

\section{Experimental probes of the current-phase relation  near $\theta_0=45^\circ$}
\label{sec:exp}
\begin{figure}[h]
	\includegraphics[width=\linewidth]{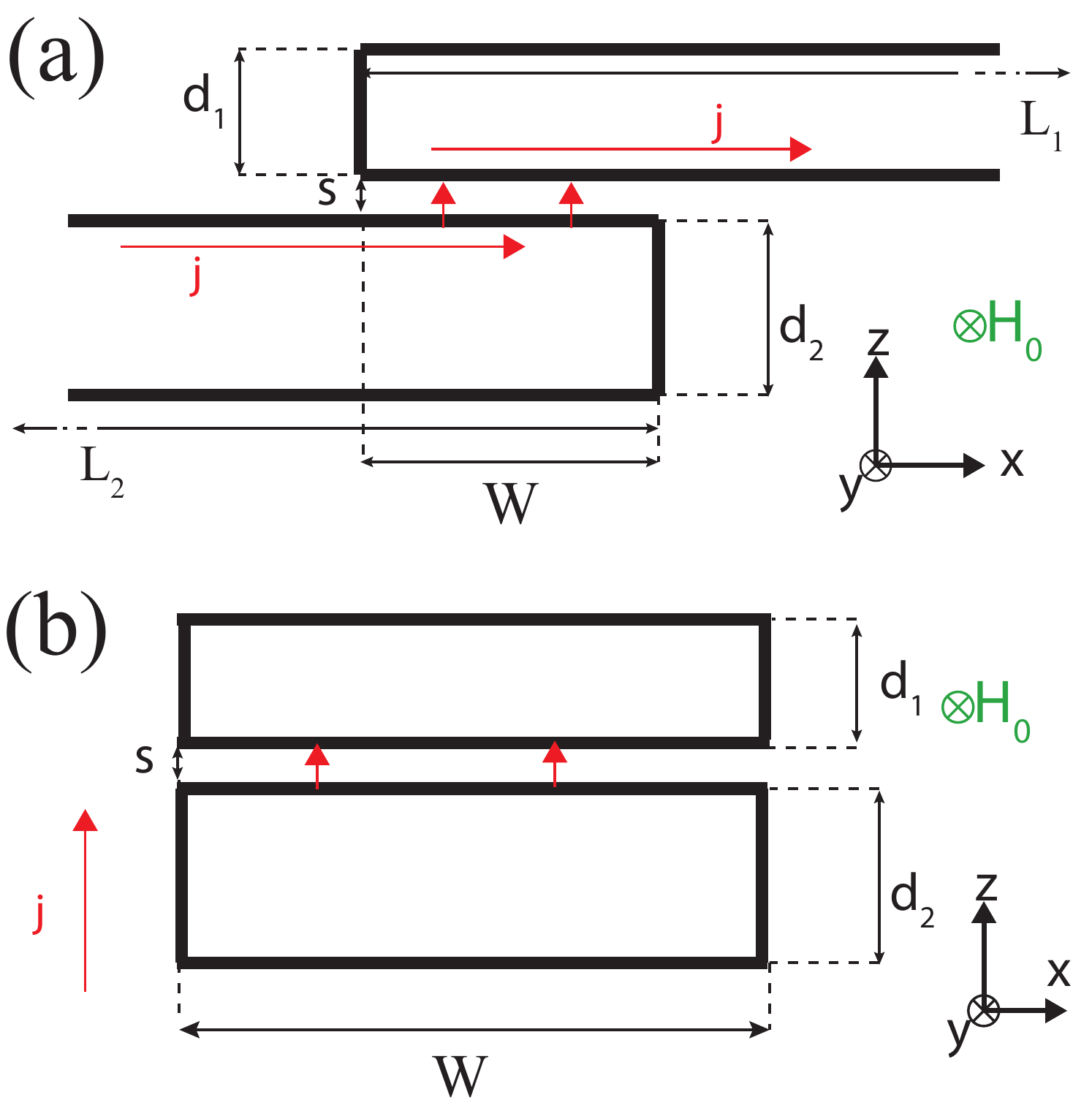}
	\caption{
	{\bf The two junction geometries considered.} 
	The in-line geometry is shown in (a) \cite{owen1967} and the vertical geometry in (b). The current (red arrows) is injected along the $x$-axes in geometry (a) and along the $z$-axes in geometry (b) . In both cases, the magnetic field is applied along the $y$ axis; length of the junctions along $y$ is denoted $D$.
	}
	\label{fig:geom}
\end{figure}

In the previous sections we have discussed the qualitative features of the dependence of the critical current in twisted $d$-wave superconducting interfaces on the twist angle  and temperature. We've established that many peculiar effects can be attributed to the lowest-order tunneling in Eq.~\eqref{eq:i2}. However, near $\theta_0=45^\circ$, the higher-order processes in Eq.~\eqref{eq:i4} of Cooper pair cotunneling  start to dominate, changing the CPR to include the second $\sin(2\varphi)$ harmonic. Here, we discuss how the CPR can be measured experimentally near $\theta_0=45^\circ$. In particular, we will address the behavior of the twist junctions in magnetic field, which results in a coordinate dependence of the phase difference across the interface $\varphi\to\varphi(x)$. We also focus on two distinct device geometries in Fig.~\ref{fig:geom}, where the set up in Fig.~\ref{fig:geom} (a) is consistent with the experimental layout of Ref.~\cite{Zhao2021}.

First, we discuss the characteristic length scales relevant for a twist junction. Generally, the characteristic variation of $\varphi$ along the length of the junction is given by the Josephson length~\cite{baronepaterno} 
\begin{equation}
  \lambda_J\sim 1/\sqrt{j_c}
\end{equation}
where $j_c$ is the critical current density. Near $\theta_0=45^\circ$, as discussed above in Secs.~\ref{sec:iccoher} and \ref{sec:disorder}, the CPR contains two sinusoidal harmonics: $\sin\varphi$ and $\sin2\varphi$, e.g. see Eqs.~\eqref{eq:ictclean} and \eqref{eq:ict4disor1}. The first harmonic corresponds to the tunneling of Cooper pairs that is required to vanish at 45$^\circ$ due the $d$-wave nature of the superconductors, while the second harmonic describes a higher-order process: co-tunneling of Cooper pairs. Correspondingly, we introduce two Josephson lengths $\lambda_{J1}(\theta_0,T)\sim 1/\sqrt{|j_c^1(\theta_0,T)|}$
and
$\lambda_{J2}(\theta_0,T)\sim 1/\sqrt{|j_c^2(\theta_0,T)|}$ (the quantitative definitions to be given below).

\begin{figure}[h]
	\includegraphics[width=\linewidth]{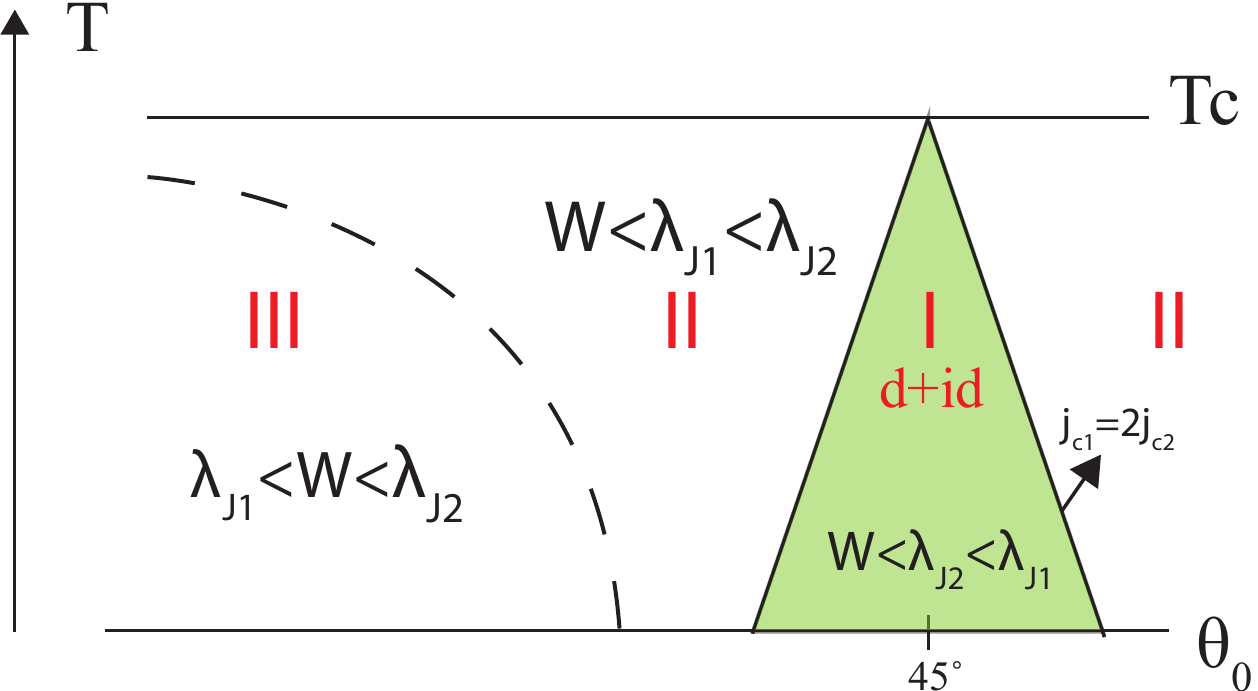}
	\caption{{\bf Qualitative phase diagram and the corresponding Josephson length $\lambda_J$  regimes for the twist junction}. The smallest length scale determines the character of spatial variations of the supercurrent across the junction: in regime $III$ the current is confined to within $\lambda_{J1}$ of the junction edges \cite{owen1967,baronepaterno}, while in $I,II$ the current is evenly distributed along the junction length. In regimes $III,II$ the first harmonic of the CPR dominates whereas  region $I$ is controlled by the second harmonic in the CPR. The boundary between $II$ and $I$ occurs  at the time-reversal symmetry breaking transition into the topological $d+id$ superconducting phase \cite{sigrist1998,can2020hightemperature}.}
	\label{fig:josregime}
\end{figure}

Denoting $W$ as the relevant linear junction size (e.g. width in the direction perpendicular to the applied field), we find three qualitative regimes, each dominated by the shortest length scale. We assume $j_c^2\ll j_c^1(\theta=0)$ due to the smallness of the interlayer tunneling at the interface and $\lambda_{J1}(\theta_0=0,T=0)<W$. Given the result in Sec. \ref{sec:iccoher}, \ref{sec:disorder}, one can then identify the position of these regimes in the $T-\theta_0$ phase diagram, as shown in Fig. \ref{fig:josregime}.

Before we move on to the magnetic field effects, it is important to remark that in region $III$ of Fig.~\ref{fig:josregime}, the device geometry will affect the superconducting properties of the junction \cite{barone1975}, in particular the value of $I_c$.  
To clarify this we consider the two device geometries depicted in Fig.~\ref{fig:geom}.
For the vertical geometry in Fig \ref{fig:geom} (b), the critical current is given by $j_c D W$; however, for an in-line geometry shown in Fig. \ref{fig:geom} (a), the critical current is equal to $4 D \lambda_J j_c\sim \sqrt{j_c}$ \cite{owen1967,barone1975}. In the latter case, the critical current is independent of the junction width $W$ and flows mostly along the junctions edges. In this case, the experimentally observed temperature and twist angle dependence of the critical current has to be compared with $\sqrt{I_c}$ from Sec. \ref{sec:iccoher},\ref{sec:disorder}, rather than that with $I_c$.

\subsection{Twist junctions in a parallel magnetic field}
We consider the Josephson effect  at the interface between two flakes of length $L_{1}$ and $L_{2}$ and thicknesses $d_{1}$ and $d_2$ with an overlap of length $W$ as depicted in Fig.~\ref{fig:geom}. We take both flakes as well as the overlap region to be of rectangular shape for simplicity; while the deviations from rectangular cross-section can affect the critical current oscillations in magnetic field \cite{baronepaterno}, they do so mostly for fields larger than the first Fraunhofer pattern zero and thus not our main focus here.

The discrete layered structure of cuprates can play an important role for magnetic field effects \cite{glazman1992}. The relevant length scale for the field variation is $s \frac{\lambda_c}{\lambda_{ab}} \lesssim \mu$m (where $s\approx 1.5$ nm is the interlayer spacing and $\frac{\lambda_c}{\lambda_{ab}}\lesssim10^3$  \cite{latyshev1996,enriquez2001}
is the ratio of the penetration depths along the $c$ axes and $ab$ plane); the overlap regions in the experiments are generally longer than that ($W\sim10\mu$m) resulting in a length-independent characteristic field \cite{fistul1994} $H_0^{FG}=\frac{\Phi_0}{\pi^2 s^2\lambda_c/\lambda_{ab}}\gtrsim 0.1$ T \cite{latyshev1996}. The characteristic fields observed for the Fraunhofer patterns near $\theta_0=45^\circ$ are  less than 100 Gauss (0.01 T) \cite{Zhao2021}. Even at the lowest fields, due to the low $H_{c1}$ values in cuprates \cite{enriquez2001}, vortices may enter the flakes, creating additional phase distortions at the junction. Note that the flakes used in the experiments are typically thinner than $\lambda_{ab}\sim 0.2\mu$m \cite{lee1996,enriquez2001} by a factor of order $2-4$, which can result in a somewhat enlarged $H_{c1}$. In the derivation below, we will ignore the presence of vortices in flakes in proximity to the junction, corresponding to sufficiently low fields i.e $H_0\lesssim H_{vort}\sim \frac{\Phi_0}{W d_{1,2}}$. For typical $W\sim10\mu$m and $d\sim0.05-0.1\mu$m, $H_{vort}$ is between $20$ and $40$ Gauss.

Consequently, limiting our considerations to sufficiently low fields to ignore the layered structure of the flakes and vortices, we can use the London equations inside the flakes to describe the screening of the magnetic field by the superconducting flakes. Note that at the interface between the two flakes (the twist junction), the phase difference can be large and this will be taken into account below.

Inside a single rectangular flake of size $d\times L$, taking the coordinate origin in its center, the London equations of the magnetic field $H(x,z)$ take the form: 
\begin{equation}
\begin{gathered}
\lambda_c^2 \frac{\partial^2 H}{\partial x^2}
+
\lambda_{ab}^2 \frac{\partial^2 H}{\partial z^2}
=
H,
\\
H|_{z=+(-)d/2} = H_0; H|_{x=\pm L/2} = H_0;
\\
H|_{z=-(+)d/2} =H_0+ H_j(x);
\end{gathered}
\label{eq:londeq}
\end{equation}
where $H_j(x)$ is the magnetic field inside the twist junction. The signs for the boundary condition along $z$ is for the case when the junction is at the bottom (top) of the flake.

The bulk of the flakes produces a Meissner effect in magnetic field, generating screening currents,  that flow through the junction affecting the phase difference across it. 
\begin{equation}
\frac{\partial H}{\partial z}
=
-\frac{4\pi}{c} j_x = 
\frac{1}{2\pi \lambda_{ab}^2}
\left[\frac{\partial \Phi}{\partial x}(x,z)+\frac{2\pi}{\Phi_0} A_x(x,z)\right],
\end{equation}
where $\Phi$ is the phase of the superconducting order parameter. Subtracting these equations at the top and the botttom of the interface (and assuming the interface thickness $s$ to be much smaller than the field variation length scale):
\begin{equation}
\left.\frac{\partial H}{\partial z}\right|_{top}(x)
-
\left.\frac{\partial H}{\partial z}\right|_{bottom}(x)
=
\frac{1}{2\pi \lambda_{ab}^2}
\left[\frac{\partial \varphi}{\partial x}(x)
+
\frac{2\pi s}{\Phi_0} H(x)\right],
\label{eq:phix}
\end{equation}
where $\varphi(x) = \Phi_{top}(x)-\Phi_{bottom}(x)+ \frac{2\pi s}{\Phi_0} A_z(x,z)$ is the gauge-invariant phase difference across the junction and $\Phi_0=\pi\hbar c/|e|$ is the flux quantum.

\subsubsection{Fraunhofer Patterns close to $\theta_0=45^\circ$}
\label{sec:fraunhofer}

For a weak junction, we can ignore the fields generated by the Josephson current, such that $H_j(x)=0$ in Eq. \eqref{eq:londeq}.
 
For $L\gg \lambda_c$ the solution away from the edges $L/2-|x|\gg\lambda_c$ can be taken as $x$-independent and has the form \begin{equation}
H_{L\to\infty}(z)=H_0 \frac{\cosh \frac{z}{\lambda_{ab}}} {\cosh \frac{d}{2 \lambda_{ab}}}.
\end{equation}
For the full problem \eqref{eq:londeq}, we use the variable-separation ansatz as described in Appendix~\ref{sec:london}.

Using the expression for $H(x,z)$ given in Appendix~\ref{sec:london} one can evaluate $\varphi(x)$ directly using Eq.~\eqref{eq:phix} (note that the flakes in the in-line geometry, Fig. \ref{fig:geom} (a), are shifted along $x$):
\begin{equation}
\begin{gathered}
-\varphi(x) = 
C
+
\frac{2\pi  H_0 s x}{\Phi_0}
-
\\
\int_0^x dx'
2\pi \lambda_{ab}^2
\left(
\left.\frac{\partial H}{\partial z}\right|_{top}(x')
-
\left.\frac{\partial H}{\partial z}\right|_{bottom}(x')
\right),
\end{gathered}
\label{eq:varphix}
\end{equation}
where $0<x<W$ and $C$ is an arbitrary dimensionless constant. The critical current across the junction is given by maximizing  over the constant $C$:
\begin{equation}
I_c(H_0, \theta_0) = \max_{C} \left\{ D\int_0^W dx j_c^1(\theta_0) \sin(\varphi) - j_c^2 \sin(2\varphi)\right\},
\label{eq:ic}
\end{equation}
where $D$ is the width of the overlap region (i.e. $D W$ is the junction area). At $\theta=45^\circ$, the first harmonic contribution to the critical current density $j_c^1(\theta_0)$ is required to vanish by symmetry \cite{klemm2005}, and has an approximately linear dependence on $\theta-45^\circ$ close to it (consistent with the lowest-harmonic  $j_c^1(\theta_0) = j_c^1(0)\cos 2\theta_0$ twist angle dependence). At the same time, $j_c^2$ does not vanish at $\theta_0=45^\circ$ and can be approximated by a constant close to it.

For an order of magnitude estimate it is convenient to use the average value of \eqref{eq:dhdz} over $x$ rather then the full $x$-dependent function. The averaging is a good approximation when $\frac{\pi W \lambda_{ab}}{d \lambda_c}\ll1$. Furthermore, the relevant dimensionless parameter for the sum $\frac{\pi\lambda_{ab}}{d}$ can be taken much larger than $1$ as the flake's thicknesses are below 100 nm, while $\lambda_{ab}\sim 0.2\mu$m \cite{lee1996,enriquez2001}. On the other hand, as is evident from \eqref{eq:dhdz}, for $\frac{\pi W \lambda_{ab}}{d \lambda_c}\gg1$, the inhomogeneities are confined to a region much smaller than the junction length $W$ and can be neglected. As is shown below, same is true if an average over $x$ is taken. We will further assume that $\frac{\pi L \lambda_{ab}}{d \lambda_c}\gg1$ for in-line geometry.

\begin{figure}[h]
	\begin{center}
		\includegraphics[width=\linewidth]{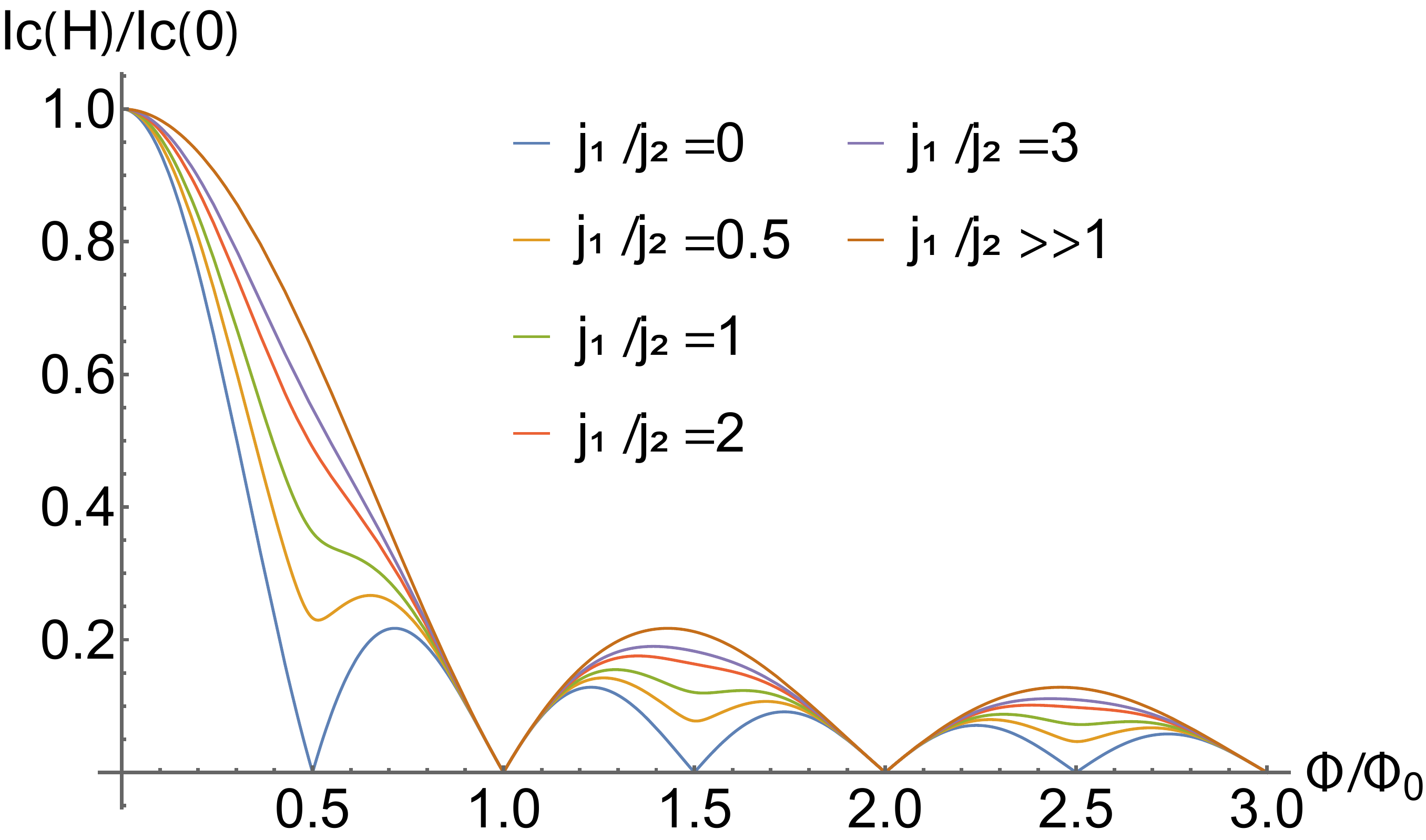}
	\end{center}
	\caption{
	{\bf Critical current versus flux displaying a Fraunhofer pattern changing its period due to the second harmonic.}
	Dependence of the critical current on the flux threading the effective junction are $\Phi=H_0 W d_{\mathrm{eff}}$ for different ratios of the first and second-harmonic critical currents. The topological transition occurs at $j_c^1=2j_c^2$.}
	\label{fig:Fr}
\end{figure}

The dependence $\varphi(x)$ is then given by:
\begin{equation}
-\varphi(x)\approx \frac{2 \pi H_0 d_{\mathrm{eff}} x}{\Phi_0}+C,
\label{eq:phixeff}
\end{equation}
where
\begin{equation}
\begin{gathered}
d_{\mathrm{eff}}
\equiv
s+
\frac{\lambda_{ab}^2 }{W}\int_0^W d x
\left(
\left.\frac{\partial H}{\partial z}\right|_{top}(x)
-
\left.\frac{\partial H}{\partial z}\right|_{bottom}(x)
\right)
\\
\approx
s
+
\sum_{i=1,2}
\lambda_{ab}\tanh \frac{d_i}{2 \lambda_{ab}}
+
\delta d_{i,\mathrm{edge}},
\end{gathered}
\label{eq:approx}
\end{equation}
where $\delta d_{i,\mathrm{edge}}$ depends on the device geometry. For the in-line  device geometry in Fig.~\ref{fig:geom}(a) we have
\begin{equation}
\begin{gathered}
    \delta d_{i,\mathrm{edge}}^{\mathrm{in-line}}
\approx
-
\frac{4 d_i}{\pi^2}
\sum_{n=0}^\infty
\frac{1}{(2n+1)^3}
\frac{1-
\exp\left(-(2n+1)\frac{W\pi\lambda_{ab}}{d_i\lambda_c}\right)
}
{\frac{W\pi\lambda_{ab}}{d_i\lambda_c}},
\end{gathered}
\end{equation}
whereas for the vertical device geometry in Fig.~\ref{fig:geom}(b) we obtain
\begin{equation}
\begin{gathered}
\delta d_{i,\mathrm{edge}}^{\mathrm{vertical}}
\approx
-
\frac{4 d_i}{\pi^2}
\sum_{n=0}^\infty
\frac{1}{(2n+1)^3}
\frac{\tanh\left((2n+1)\frac{W\pi\lambda_{ab}}{2d_i\lambda_c}\right)
}
{\frac{W\pi\lambda_{ab}}{2 d_i\lambda_c}}.
\end{gathered}
\end{equation}
Importantly, this result depends on two dimensionless parameters: $\frac{d_i}{2 \lambda_{ab}}$ and $\frac{W\pi\lambda_{ab}}{d\lambda_c}$. For a purely first-harmonic current-phase relation (i.e. $j_c^2=0$ in \eqref{eq:ic}) one obtains then the conventional Fraunhofer pattern, with the first zero being at a field:
\begin{equation}
H^{(1)}_0 = \frac{\Phi_0}{W d_{\mathrm{eff}}},
\label{eq:expdeff1}
\end{equation}
Which allows to extract the value of $d_{\mathrm{eff}}$ from the experimentally observed Fraunhofer pattern. Note that in the opposite case $j_c^1=0$ (i.e. at $\theta_0=45^\circ$) the first zero in the pattern occurs at 
\begin{equation}
H^{(2)}_0 = \frac{\Phi_0}{2 W d_{\mathrm{eff}}},
\label{eq:expdeff2}
\end{equation}
which implies a twice smaller $d_{\mathrm{eff}}$ value for the same Fraunhofer pattern. When both $j_c^1$ and $j_c^2$ are nonzero, the dependence $I_c(H)$ obtained from Eq. \eqref{eq:ic} interpolates between the two limits as is shown in Fig. \ref{fig:Fr}

\begin{figure}[h]
	\begin{center}
		\includegraphics[width=\linewidth]{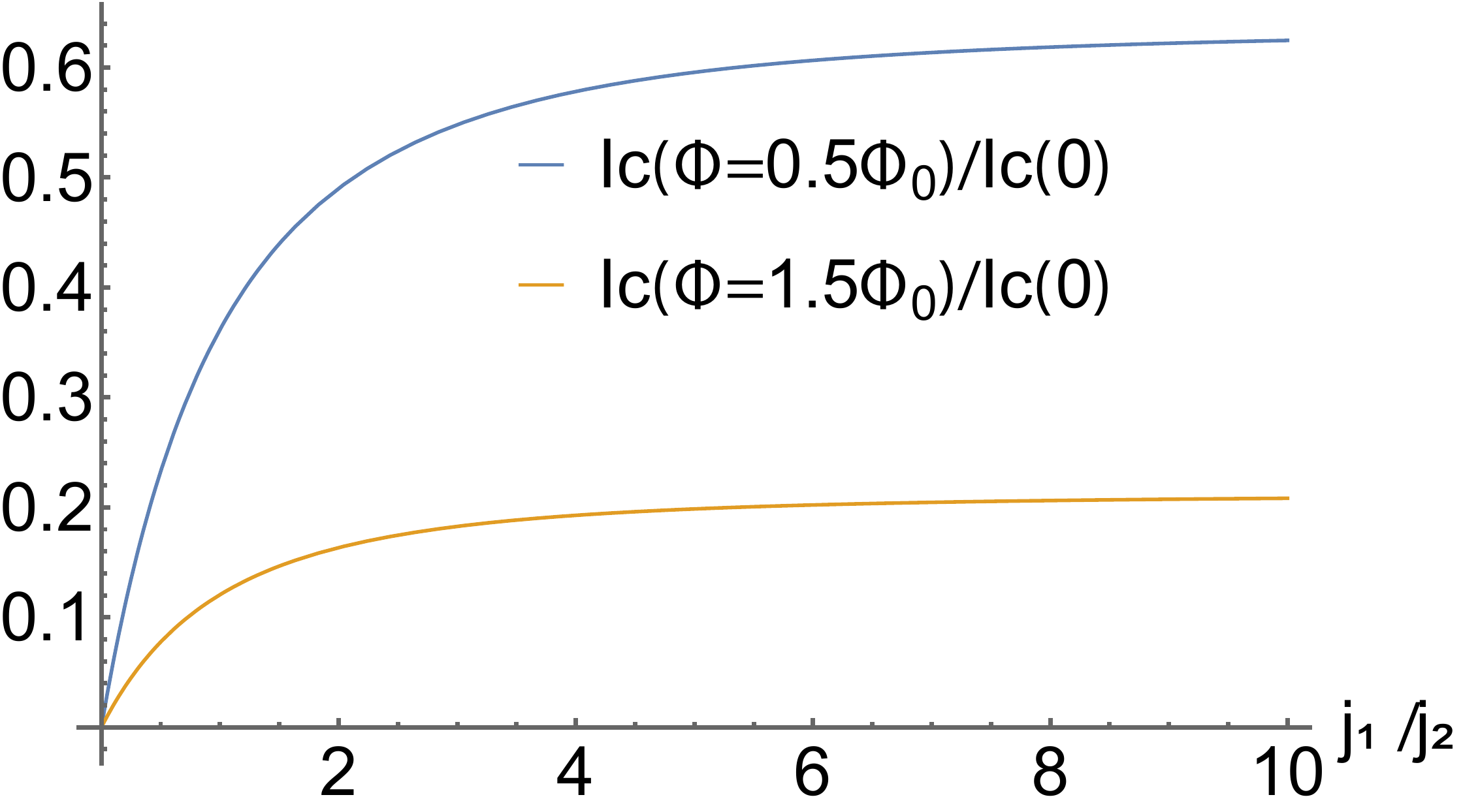}
	\end{center}
	\caption{
	{\bf Lifting the nodes of the second harmonic Fraunhofer pattern.}
	Dependence of the critical current at half-integer flux values on the ratio between $j_c^1$ and $j_c^2$.}
	\label{fig:Fr2}
\end{figure}

Particularly sensitive are the odd-numbered zeroes of the second-harmonic pattern, that are visibly lifted by a nonzero $j_c^1$. In Fig. \ref{fig:Fr2}, this lifting is demonstrated more quantitatively. Indeed, for $j_c^1\gtrsim2j_c^2$ the values are almost indistinguishable from those at $j_c^2=0$.

Note that $d_{\mathrm{eff}}$ can be temperature-dependent via the penetration depths $\lambda_{ab/c}(T)$, which have to diverge at $T_c$. However, the dependence on $\frac{d_i}{2 \lambda_{ab}(T)}$ can be  neglected as can be seen from using the lower bound for $\lambda_{ab}(0)\gtrsim0.21\mu$m \cite{enriquez2001} and thickness $d\lesssim100$ nm, we find that $0.49 d_i<d_i\frac{\lambda_{ab}(0)}{d_i}\tanh \frac{d_i}{2 \lambda_{ab}(0)}\leq d_i/2$, i.e. a variation below $1\%$, much less than the one observed in the experiment \cite{Zhao2021}. Thus, we take $\lambda_{ab}\tanh \frac{d_i}{2 \lambda_{ab}} \to d_i/2$ in \eqref{eq:approx}.

While $d_{\mathrm{eff}}$ can depend on the device geometry and smoothly on temperature, Fig. \ref{fig:josregime} suggests that a robust evidence of the second harmonic in CPR can be obtained for a single device with a twist angle close to $\theta_0=45^\circ$. In particular, even if the cotunneling contribution, Eq. \eqref{eq:i4}, leading to second harmonic in CPR, is dominant at low $T$, it has to become negligible with respect to the usual tunneling current close to $T_c$ (i.e. a transition from $I$ to $II$ occurs on heating). Thus, lifting of the odd-numbered nodes in the Fraunhofer pattern on heating  represents an unambiguous evidence that first and second harmonic coexist in CPR.

\subsubsection{Away from $\theta_0=45^\circ$: crossover to long-junction limit}

Away from $\theta_0=45^\circ$, the critical current density of the twist junction grows strongly and one can not ignore the effect  of this current on magnetic field anymore.

To start with a concrete but simple example, we first discuss the case of two monolayers of a nodal superconductor in magnetic field (i.e. both flakes in Fig. \ref{fig:geom} being monolayers). The current in the monolayer flowing along $x$ is given by \cite{glazman1992}:
\begin{equation}
    j_x(x,z) =-\frac{c \Phi_0 s'}{ 8\pi^2 \lambda_{ab}^2}
    \left(\partial_x \Phi_{1,2}(x)+\frac{2\pi}{\Phi_0}A_x\right) \delta(z-z_{1,2}),
\end{equation}
where $s'$ is the monolayer thickness and $z_{1,2}$ - its coordinate along $z$ (where $z_1-z_2=s$), $\Phi_0=\pi\hbar c/|e|$ is the flux quantum. We denote the magnetic field between the monolayers as $H_0+H_j(x)$ (outside it is equal to $H_0$). Integrating the Maxwell's equation $-\frac{\partial H}{\partial z} = \frac{4\pi}{c}j_x$ across each flake and subtracting the results we get:
\begin{equation}
\begin{gathered}
      H_j(x) = - \frac{s' \Phi_0}{4\pi\lambda_{ab}^2} 
      \left(
      \partial_x(\Phi_1-\Phi_2)
      \right.
      \\
      \left.+\frac{2\pi}{\Phi_0} [A_x(z=z_1)-A_x(z=z_2)]
      \right).
\end{gathered}
\end{equation}
Assuming the magnetic field variations to occur at a scale much larger than $s$ we can further bring this equation to the form:
\begin{equation}
H_j(x) \approx -\frac{\frac{\Phi_0}{2\pi} \partial_x \varphi}{\frac{2\lambda_{ab}^2}{s'}+s}
-\frac{H_0\frac{s's}{2\lambda_{ab}^2}}{1+\frac{s's}{2\lambda_{ab}^2}}
\end{equation}
where we introduced the gauge-invariant phase difference across the junction:
\begin{equation}
    \varphi(x) = \Phi_1(x)-\Phi_2(x)+\frac{2\pi}{\Phi_0}\int_{z_1}^{z_2} A_z dz.
\end{equation}
Finally we can get a closed equation for $\varphi(x)$ in the case of twisted monolayers (ml) using $\frac{\partial H}{\partial x} = -\frac{4\pi}{c} (j_c^{(1)} \sin (\varphi)+j_c^{(2)} \sin (2\varphi))$:
\begin{equation}
    \begin{gathered}
       \partial_{xx} \varphi
    =
    \frac{\sin(\varphi)}{\lambda_{J1,\mathrm{ml}}^2}
    +\frac{\sin(2\varphi)}{\lambda_{J2,\mathrm{ml}}^2},
    \\
    \lambda_{J1(2),\mathrm{ml}}^2 = \frac{c |\Phi_0|}{8\pi^2 j_c^{(1(2))}\left(\frac{2\lambda_{ab}^2}{s'}+s\right)}.
    \end{gathered}
    \label{eq:longjuncbilayer}
\end{equation}
For $s'=s$, $\lambda_{ab}\gg s$ and $j_c^{(1)} = \frac{c|\Phi_0|}{8\pi^2 s \lambda_c^2}$ we recover the known result $\lambda_{J1,\mathrm{ml}}^2 = s^2 \gamma^2/2$ \cite{glazman1992}, where $\gamma= \lambda_c/\lambda_{ab}$. The boundary conditions for this equation are determined by the external field and the current in the in-plane geometry.

As has been shown above, the Josephson length for twisted monolayers $\lambda_{J1,\mathrm{ml}}$ decreases away from $\theta_0=45^\circ$ rapidly and hence the second harmonic term in \eqref{eq:longjuncbilayer} can be neglected, reducing it to the usual equation describing a long Josephson junction \cite{owen1967}. The solution of this problem is well-known and we shall not reproduce it here: for $\lambda_{J1,ml}\lesssim W$ while the critical current is still suppressed by field, no clear Fraunhofer pattern is expected: in particular, $I_c(H)$ exhibits no zeroes at finite fields \cite{owen1967}.

We now can discuss to the case of finite-thickness flakes. To allow for analytical closed-form expression we will focus on the vertical junction geometry, Fig. \ref{fig:geom} (b). To include the effects of the junction self-field we will follow an approach similar to Ref. \onlinecite{alfimov1995}. In particular, we first solve the equation \eqref{eq:londeq} for an arbitrary function $H_j(x)$ and then reexpress the magnetic field inside the junction via the phase difference $\varphi(x)$ using Eq. \eqref{eq:phix}. Finally, using $\frac{\partial H}{\partial x} = j_c^{(1)} \sin(\varphi)$ (we neglect the second harmonic here as $\theta_0$ is far fro $\theta_0=45^{\circ}$) and Appendix~\ref{sec:london} we obtain:
\begin{equation}
    \int_{-W/2}^{W/2}K(x-x') \partial_{x'x'} \varphi(x')
    = \frac{8\pi^2  j_c^{(1)}}{c \Phi_0} \sin(\varphi_0(x)+\delta\varphi(x) ),
    \label{eq:josnonloc}
\end{equation}
where the Kernel is given by 
  \begin{equation}
   K(x,x') = \frac{1}{W}
    \sum_{n>0} \frac{
    \cos [k_n(x+L/2)]
    \cos [k_n(x'+L/2)]
    }
    {s+\sum_{i=1,2} \lambda_{ab} \sqrt{1+k_n^2\lambda_c^2}
\coth\frac{d \sqrt{1+k_n^2\lambda_c^2}}{\lambda_{ab}}}.   
  \end{equation}
The expression \eqref{eq:josnonloc} can then be analyzed in several limiting cases. In particular, 
\begin{equation}
    K(x,x')=
    \begin{cases}
\frac{\delta(x-x')}{s}&s\gg\lambda_{ab}, \\
\frac{\delta(x-x')}{s+\sum_i\frac{\lambda_{ab}^2}{d_i}}&d_i\ll\lambda_{ab}.
    \end{cases}
\end{equation}
For both of this cases, the resulting Josephson length is given by:
\begin{equation}
    \lambda_{J,fl}^2 = \frac{c \Phi_0} {8\pi^2 j_c^{(1)}\left(s+\sum_i\frac{\lambda_{ab}^2}{d_i}\right)}.
    \label{eq:ljosnonloc}
\end{equation}
Importantly, the reduction of the effective thickness, evident in \eqref{eq:phixeff} does not show up here in the same way as for conventional Josephson junctions. Expression Eq. \eqref{eq:ljosnonloc} implies the limit on the critical current density for the observation of the Fraunhofer pattern:
\begin{equation}
    j_c^{(1)} <\frac{c \Phi_0} {8\pi^2 W^2\left(s+\sum_i\frac{\lambda_{ab}^2}{d_i}\right)}.
    \label{eq:jcfraun}
\end{equation}

For $d\gg\lambda_{ab}$ the problem becomes manifestly non-local; however, equation \eqref{eq:ljosnonloc} can be used as an order of magnitude estimate in this case. For $W\ll\lambda_c$, on the other hand, the relevant length scale is of the order $ \lambda_{J,fl}^2 \sim \frac{c \Phi_0} {j_c^{(1)}\lambda_{ab}\lambda_c/W}$. The critical value of the Josephson current (for the observation of the Fraunhofer pattern) is of the order $\frac{c \Phi_0} {W\lambda_{ab}\lambda_c}$, smaller than the one in Eq. \eqref{eq:jcfraun} (assuming $s\ll\lambda_{ab}^2/d_i$) by $\sim W\lambda_{ab}/(d\lambda_c)$.

Overall, we have shown that for twisted flakes of $d$-wave superconductors, a clear Fraunhofer pattern appears close to $\theta_0=45^\circ$ twist, with features indicating the presence of a second harmonic in the CPR. Away from $\theta_0=45^\circ$, the Fraunhofer pattern will be smeared progressively due to the importance of the magnetic field generated by the twist junction itself.

\section{Conclusions}
\label{sec:conc}
In this Article, we have studied the Josephson effect in twisted bilayers of nodal superconductors and analyzed experimental setups that can be used to measure the current-phase characteristics of these devices.

We have demonstrated that the temperature dependence of the critical current is quite generally expected to have a nonmonotonic form due to the negative contribution of the near-nodal region in momentum space. The critical current is strongly suppressed on increasing the twist angle, with the precise form determined by the Fermi surface geometry, momentum dependence of the tunneling and roughness of the interface. At $\theta_0=45^\circ$, the critical current reaches a nonzero minimum due to the Cooper pair cotunneling processes.

Dependence of the critical current on magnetic field has been studied including the effects of the sample geometry and for finite-thickness flakes forming the junction. At $\theta_0=45^\circ$, we have demonstrated that a clear Fraunhofer pattern with halved period should be observed; at elevated temperatures or away from $\theta_0=45^\circ$ the odd-numbered zeroes are lifted suggesting a robust signature of the coexistence of tunneling and cotunneling of Cooper pairs. Further away from $\theta_0=45^\circ$ the Fraunhofer pattern is shown to vanish due to self-field effects and we have calculated the critical current density for this crossover.

Finally, our results  reproduce the main features observed in the recent experiments on twist junctions of high-T$_c$ cuprates \cite{Zhao2021}. In summary, this inlcudes the $I_c\sim\cos(2\theta_0)$ dependence of the critical current, its nonmonotonic temperature dependence with a maximum at nonzero temperature, and the emergence of a Fraunhofer-like dependence on magnetic field close to $\theta_0=45^\circ$.

\section*{Acknowledgments}
P.A.V.\ is supported by a Rutgers Center for Material Theory Postdoctoral Fellowship and
J.H.P.\ is partially supported by the Air Force Office of Scientific Research under Grant No.~FA9550-20-1-0136, NSF CAREER Grant No. DMR-1941569, and  the Alfred P. Sloan Foundation through a Sloan Research Fellowship. 
The Flatiron Institute is a
division of the Simons Foundation. P.A.V.\ and J.H.P.\ acknowledge the Aspen Center for Physics where part of this work was performed, which is supported by National Science Foundation grant PHY-1607611. This work was partially supported by a grant from the Simons Foundation (P.A.V.). N.P. acknowledges the Deutsche Forschungsgemeinschaft (DFG452128813) for partial support with the project.
P.K. acknowledge the support from the NSF (DMR-1809188) and S.Y.F.Z and X.C acknowledge the support from NSF (DMR-1922172).

\appendix
\section{Evaluating the tunneling contribution in the presence of interface roughness}
\label{sec:appendixA}

We can now perform the angular integration in Eq. (\ref{eq:i2}). To do this, we rewrite \eqref{eq:i2} using Fourier series:
\begin{equation}
	\begin{gathered}
	I^{(2)}(\varphi,T,\theta_0)
	=
	\\
   	\frac{4e\sin \varphi}{\hbar} T\sum_{\varepsilon_n}
    \int \frac{kdk}{(2\pi)^2} \frac{k'dk'}{(2\pi)^2} 
    \frac{t_0^2 }{2 \pi \sigma^2} e^{-\frac{(k - k')^2}{2\sigma^2}}
    I^{(2)}_{ang}(\varepsilon_n,\theta_0,k,k'),
	\end{gathered}
	\label{eq:i2incoh}
\end{equation} 
where
\begin{equation}
	\begin{gathered}
    I^{(2)}_{ang}(\varepsilon_n,\theta_0,k,k')
    =
    \int
	d \theta  d \theta' 
    e^{-\frac{(\theta - \theta')}{2\tilde{\sigma}^2}}
	\sum_n f_n(k) \cos 2 n \theta
	\\
	\cdot
	 \sum_m f_m(k') \cos 2 m (\theta'+\theta_0)
	 =
	 \\
	 =
	 \pi \sqrt{2\pi\tilde{\sigma}^2}
	 \sum_m f_{m}(k)f_{m}(k') 
	 e^{-2\tilde{\sigma}^2 m^2}
	 \cos 2 m \theta_0,
	\end{gathered}
	\label{eq:i2angav}
\end{equation} 
where we assumed $\tilde{\sigma} \ll \pi$ and
\begin{equation}
\begin{gathered}
    f_{m\neq 0}(k) = \int \frac{d\theta}{\pi} 
     \frac{\Delta(\bk)\cos(2 m \theta)}
   {\varepsilon_n^2+\xi^2(\bk)+\Delta^2(\bk)},
    \\
    f_{m=0}(k) = 0,
\end{gathered}
\label{eq:fFourier}
\end{equation} 
are the Fourier coefficients of the anomalous Green's functions. One observes already that the nonconservation of the angular component of the momentum suppresses the oscillatory behavior of $I^{(2)}(\varphi,T,\theta_0)$ as a function of $\theta_0$. Indeed, in \eqref{eq:i2angav} the high harmonics ($n \gg \frac{1}{\tilde {\sigma} }$) are strongly suppressed (the precise form depends however, on the realization of momentum smearing in the tunneling).

To make further progress analytically, we take $\xi(\bk)=\xi(k)$ and $\Delta(\bk) = \Delta(T)\cos 2\theta $ as in section \ref{sec:deltaT} and limit ourselves to the lowest terms in the Fourier series, Eq. \eqref{eq:i2angav}. This results  in $f_{0}(\xi) =  0$ and $f_{2}(\xi) =  0$ (see Eq. \eqref{eq:fFourier}) while
\begin{equation}
\begin{gathered}
    f_{1}(\xi) =  \frac{2}{\Delta(T)} \left(1- \sqrt{\frac{\varepsilon^2+\xi^2}{\varepsilon^2+\xi^2+\Delta^2(T)}}\right)
    \\
    f_{3}(\xi) =  \frac{2}{\Delta^3(T)} \left(-4(\varepsilon^2+\xi^2)-\Delta^2(T)+
    \right.
    \\
    \left.
    +(4(\varepsilon^2+\xi^2)+3\Delta^2(T))
    \sqrt{\frac{\varepsilon^2+\xi^2}{\varepsilon^2+\xi^2+\Delta^2(T)}}\right).
\end{gathered}
\end{equation}

For smooth inhomogeneity we assume $\hbar v_F \sigma \ll \sqrt{\Delta^2(T)+(\pi T)^2}$ that can be valid at all $T$. 
We can further simplify the result by taking the limit $\sigma\to 0$ in the $k,k'$ integral in (see Eq.~\eqref{eq:i2incoh}) to obtain
\begin{equation}
\begin{gathered}
I^{(2)}_{\mathrm{SI}}(T,\varphi,\theta_0) \approx
\frac{e t_0^2 \sin \varphi k_F} {4 \pi^3 \hbar^2 v_F}
(i^{(2)}_1(T)\cos(2\theta_0)e^{-2\tilde{
\sigma}^2}
\\
+i_3(T)\cos(6\theta_0)e^{-18\tilde{\sigma}^2}),
\\
i^{(2)}_1(T) = T \sum_{\varepsilon_n}  \int d \xi 
f_1^2(\xi)
\approx
\begin{cases}
2(\log 4 -1) &T\to 0\\
0.1 \frac{\Delta^2(T)}{T_c^2} &T\to T_c
\end{cases}
\\
i^{(2)}_3(T) = T \sum_{\varepsilon_n}  \int d \xi 
f_3^2(\xi)
\approx
\begin{cases}
2(\log 4 -4/3) &T\to 0\\
4\cdot10^{-5}\frac{\Delta^6(T)}{ T_c^6} &T\to T_c
\end{cases}
\end{gathered}
\label{eq:ictclean2}
\end{equation}

In the opposite limit of rough inhomogeneity $\hbar v_F \sigma \gg \sqrt{\Delta^2(T)+(\pi T)^2}$ 
we obtain a result that is consistent with the more usual Ambegaokar-Baratoff \cite{ABerr} like expression:
\begin{equation}
\begin{gathered}
I^{(2)}_{\mathrm{SD}}(T,\varphi,\theta_0) \approx
\frac{e t_0^2 \sin \varphi k_F} {4 \pi^3 \hbar^2 v_F}
\frac{1}{\sqrt{2\pi}\hbar v_F \sigma}
\\
\tilde{i}^{(2)}_1(T)\cos(2\theta_0)e^{-2\tilde{\sigma}^2}
+\tilde{i}^{(2)}_3(T)\cos(6\theta_0)e^{-18\tilde{\sigma}^2}),
\\
\tilde{i}^{(2)}_1(T) = T \sum_{\varepsilon_n}  \left(\int d \xi 
f_1(\xi)\right)^2
\approx
\begin{cases}
6.035 \Delta(T) &T\to 0\\
\frac{\pi^2}{4} \frac{\Delta^2(T)}{T_c} &T\to T_c
\end{cases}
\\
\tilde{i}^{(2)}_3(T)= T \sum_{\varepsilon_n}  \left(\int d \xi 
f_3(\xi)\right)^2
\approx
\begin{cases}
0.18 \Delta(T) &T\to 0\\
3\cdot10^{-4}\frac{\Delta^2(T)}{T_c} &T\to T_c
\end{cases}
\end{gathered}
\label{eq:ictdirty-all}
\end{equation}
In both cases the $\cos(6\theta_0)$ contribution appears to be strongly suppressed numerically (in addition to the exponential suppression due to angular spread): by an almost order of magnitude at low $T$ and by several orders of magnitude close to $T_c$.

We now apply similar calculations to obtain the fourth-order tunneling contribution to the critical current $I^{(4)}$ in Eq.~\eqref{eq:i4}. By applying the same expansion in twist angle harmonics we obtain the following result for the leading twist-angle harmonics 
\begin{equation}
\begin{gathered}
   I^{(4)} = I^{(4)}_{1,c}(T) \cos 2\theta_0 \sin\varphi+I^{(4)}_{2,c}(T,\theta_0)\sin 2\varphi 
   \\
   I^{(4)}_{1,c}(T) =   - \frac{e t_0^4 k_F \sigma^2 e^{-3\tilde{\sigma}^2}}{\hbar^2 v_F (2\pi)^6} 
   i^{(4)}_1(T),
   \\
  I^{(4)}_{2,c}(T,\theta_0)
  =
   - \frac{e t_0^4 k_F \sigma^2 e^{-4\tilde{\sigma}^2}}{\hbar^2 v_F (2\pi)^7} 
   (\cos 4\theta_0+2 e^{-4\tilde{\sigma}^2})
   i^{(4)}_2(T),
   \end{gathered}
   \label{eq:ict4disor}
\end{equation}
where
\begin{equation}
    \begin{gathered}
i^{(4,\mathrm{SI})}_1(T) = T \sum_{\varepsilon_n}  \int d \xi 
f_1^2(\xi)(\varepsilon_n^2-\xi^2)g_0^2(\xi),
\\
i^{(4,\mathrm{RI})}_1(T) = \frac{T\sum_{\varepsilon_n}  \left(\int d \xi 
f_1(\xi)\right)^2\left(\int d \xi 
\varepsilon_n g_0(\xi)\right)^2}{2(\sqrt{\pi}\hbar v_F\sigma)^3},
\\
i^{(4,\mathrm{SI})}_2(T) = T \sum_{\varepsilon_n}  \int d \xi 
f_1^4(\xi),
\\
i^{(4,\mathrm{RI})}_2(T) = \frac{T\sum_{\varepsilon_n}  \left(\int d \xi 
f_1(\xi)\right)^4}{2(\sqrt{\pi}\hbar v_F\sigma)^3}.
    \end{gathered}
\end{equation}
Recall that SI and RI label the smooth and rough inhomogeneity regimes, respectively, and we have introduced
\begin{equation}
    \begin{gathered}
g_0(k) = \int \frac{d\theta}{2\pi} 
     \frac{1}
   {\varepsilon_n^2+\xi^2(\bk)+\Delta^2(\bk)}.
    \end{gathered}
\end{equation}
\section{Solution of London equation}
\label{sec:london}
\subsection{Without self-field effects}
To solve the London equation in Eq.~\eqref{eq:londeq}  we use
the variable-separation ansatz:
\begin{equation}
\begin{gathered}
H_1(x,z)= 
H_{L\to\infty}(z)
+
\\
\sum_{n>0} C_n 
\frac{\cosh\left(\frac{x}{\lambda_c}\sqrt{\left(\frac{(2n+1)\pi\lambda_{ab}}{d}\right)^2+1}\right)}
{\cosh\left(\frac{L}{2\lambda_c}\sqrt{\left(\frac{(2n+1)\pi\lambda_{ab}}{d}\right)^2+1}\right)}
\cos\frac{(2n+1)\pi z}{d},
\end{gathered}
\end{equation}
where we use $H(x,z) = H(x,-z)$ and $H|_{z=\pm d/2} = H_0$ and $H|_{z=\pm L/2} = H_0$. We denote this solution as $H_1(x,z)$ to highlight that the self-field of the twist junction has been neglected. To determine the coefficients $C_n$ we use the boundary conditions at the ends of the flake $H|_{x=\pm L/2} = H_0$:
\begin{equation}
\begin{gathered}
C_n = \frac{2H_0}{d}\int_{-d/2}^{d/2} dz
 \left(1-\frac{\cosh \frac{z}{\lambda_{ab}}} {\cosh \frac{d}{2 \lambda_{ab}}}\right)
 \cos\frac{(2n+1)\pi z}{d}=
 \\
 =
 \frac{4 H_0 (-1)^n}{(2n+1)\pi}
 -
  \frac{4 H_0 (-1)^n\frac{(2n+1)\pi\lambda_{ab}^2}{d^2}}{\left(\frac{(2n+1)\pi\lambda_{ab}}{d}\right)^2+1}
 =
 \\
 =
 \frac{4H_0(-1)^n}{(2n+1)\pi}
\frac{1}{\left(\frac{(2n+1)\pi\lambda_{ab}}{d}\right)^2+1}
\end{gathered}
\end{equation}
where we used $\int \cosh z \cos a z = \frac{a \cosh z \sin a z+ \sinh z \cos a z }{a^2+1}$. Finally, the full solution for $H_1(x,z)$ for $H_j(x)=0$ is given by:
\begin{equation}
\begin{gathered}
H_1(x,z)
=
H_0 \frac{\cosh \frac{z}{\lambda_{ab}}} {\cosh \frac{d}{2 \lambda_{ab}}}+
\\
+H_0 \sum_{n=0}^\infty  \frac{4(-1)^n}{(2n+1)\pi}
\frac{\cos\frac{(2n+1)\pi z}{d}}{\left(\frac{(2n+1)\pi\lambda_{ab}}{d}\right)^2+1}\times
\\
\times \frac{\cosh\left(\frac{x}{\lambda_c}\sqrt{\left(\frac{(2n+1)\pi\lambda_{ab}}{d}\right)^2+1}\right)}
{\cosh\left(\frac{L}{2\lambda_c}\sqrt{\left(\frac{(2n+1)\pi\lambda_{ab}}{d}\right)^2+1}\right)}
\end{gathered}
\label{eq:hxzfull}
\end{equation}
The relevant quantity that enters Maxwell's equations at the junction's edges is 
\begin{equation}
\begin{gathered}
\pm\left.\frac{\partial H_1}{\partial z}(x)\right|_{z=\pm d/2}  = 
\frac{H_0}{\lambda_{ab}} \tanh \frac{d}{2 \lambda_{ab}} 
-
\\
-
\frac{4 H_0}{d}
\sum_{n=0}^\infty
\frac{1}{\left(\frac{(2n+1)\pi\lambda_{ab}}{d}\right)^2+1}
\frac{\cosh\left(\frac{x}{\lambda_c}\sqrt{\left(\frac{(2n+1)\pi\lambda_{ab}}{d}\right)^2+1}\right)}
{\cosh\left(\frac{L}{2\lambda_c}\sqrt{\left(\frac{(2n+1)\pi\lambda_{ab}}{d}\right)^2+1}\right)}.
\end{gathered}
\label{eq:dhdz}
\end{equation}

\subsection{Including self-field effects}
We can write the solution as $H(x,z) = H_1(x,z)+\delta H(x,z)$, where $H_1(x,z)$ is given by \eqref{eq:hxzfull}. $\delta H(x,z)$ satisfies zero boundary conditions except for the surface of the junction, where is is equal to $H_j(x)$. The solution can be obtained by variable separation ansatz that yields:
\begin{equation}
    \begin{gathered}
    \delta H(x,z) = \sum_{n\neq0} h_n
   \frac{\sinh \frac{(d/2-z) \sqrt{1+\lambda_c^2 k_n^2}}{\lambda_{ab}}}
   {\sinh \frac{d \sqrt{1+\lambda_c^2 k_n^2}}{\lambda_{ab}}}
   \sin[k_n(x+W/2)]
   ,
   \\
   h_n = \frac{1}{W}\int_{-W/2}^{W/2} dx H_j(x) 
   \sin[k_n(x+W/2)]
   ,
    \end{gathered}
\end{equation}
where $k_n=\frac{\pi n}{W}$. Next, we need to express the extra field in the junction via the phase difference \eqref{eq:phix}. In particular, we can use the result of Sec. \ref{sec:fraunhofer} and include the correction due to the Josephson self-field as:
\begin{equation}
    \begin{gathered}
    \varphi(x) \to \varphi(x)+\delta \varphi(x),
    \\
    \delta \varphi(x) \equiv \sum_{n>0} \delta \varphi_n \cos\left(\frac{\pi n x}{W}+\frac{\pi n}{2}\right),
    \end{gathered}
\end{equation}
where $\varphi(x)$ is given by Eqs. (\ref{eq:varphix}, \ref{eq:phixeff}).
From Eq. \eqref{eq:phix} we find
\begin{equation}
\delta h_n = - \frac{\Phi_0}{2\pi}
\frac{k_n \delta \varphi_n}
{s+\sum_{i=1,2} \lambda_{ab} \sqrt{1+k_n^2\lambda_c^2}
\coth\frac{d \sqrt{1+k_n^2\lambda_c^2}}{\lambda_{ab}}
}
\end{equation}

\bibliography{allrefs}
\end{document}